\documentclass[11pt]{article}
\usepackage{graphicx}
\newcommand{\Tr}{\mbox{\rm Tr}}
\newcommand{\be}{\begin{equation}}
\newcommand{\ee}{\end{equation}}
\newcommand{\bea}{\begin{eqnarray}}
\newcommand{\eea}{\end{eqnarray}}
\newcommand{\nn}{\nonumber}
\newcommand{\f}[1]{\,\mbox{\raisebox{0.15ex}{\footnotesize $#1$}}\,}
\newcommand{\Lg}{\mbox{\rm Lg}}
\newcommand{\Ka}{{\!\!\;\mbox{\it\tiny K}}}
\newcommand{\Diag}[4]{\mbox{$\Delta^{(#1)}_{#2}$}}
\newcommand{\BOne}[5]{B(#3,#4)}
\newcommand{\BOneNenner}[7]{B^\prime(#3,#5)}
\newcommand{\ANenner}[2]{A^\prime(#2)}
\newcommand{\vierpi}[1]{} %%% (4\pi)^#1
\newcommand{\TFull}[9]{T_{#1,#2,#3}}
\newcommand{\TSun}[6]{S_{#1, #2}}
\newcommand{\Lren}[1]{\!\; {L}_{#1}^{\mbox{\scriptsize\em ren}} \!\: }

\begin{document}

%********************************************************************
%  TITLE PAGE
%********************************************************************

\begin{flushright}
   MZ-TH--00-31  \\
   hep-ph/0007095\\
   July 2000 \\
\end{flushright}
\vspace*{3cm}
\begin{center} {\bf \large
  \boldmath $K^0$ form factor at order $p^6$ of chiral perturbation theory
  } \vspace{10 mm} \\
  P.~Post 
  \footnote{Now at GMD - Forschungszentrum Informationstechnik GmbH, 
  Schloss Birlinghoven, D-53754 Sankt Augustin, Germany, 
  E-mail: peter.post@gmd.de} 
  and K.~Schilcher
  \footnote{Supported in part by Bundesministerium f\"ur Bildung und
  Forschung, Bonn, Germany, under Contract 05~HT9UMB~4, 
  E-mail: karl.schilcher@uni-mainz.de}
  \vspace{10 mm} \\
  {\em Institut f\"ur Physik, Johannes-Gutenberg-Universit\"at,\\
  Staudinger Weg 7, D-55099 Mainz, Germany}
\end{center}
\vspace{12 mm}
\begin{abstract}
This paper describes the calculation of the electromagnetic form
factor of the $K^0$ meson at order $p^6$ of chiral perturbation
theory which is the next-to-leading order correction to the
well-known $p^4$ result achieved by Gasser and Leutwyler.
On the one hand, at order $p^6$ the chiral expansion contains 
1- and 2-loop diagrams which are discussed in detail. 
Especially, a numerical procedure
for calculating the irreducible 2-loop graphs of the \emph{sunset}
topology is presented. On the other hand, the chiral Lagrangian
${\cal L}^{(6)}$ produces a direct coupling of the $K^0$ current with
the electromagnetic field tensor. Due to this coupling one of the
unknown parameters of ${\cal L}^{(6)}$ occurs in the contribution
to the $K^0$ charge radius.
\end{abstract}

\newpage

%********************************************************************
%  TEXT
%********************************************************************

\section{Introduction}

Chiral perturbation theory has established itself as a powerful
effective theory of low energy strong interactions. Based on the
symmetry of the underlying QCD, chiral perturbation theory produces a
systematic low-energy expansion of the observables in this regime.
Unfortunately, because of the non-renormalizability of the effective
theory, higher powers in the energy expansion require higher loop
Feynman integrals and as input an ever increasing number of
renormalization constants. The $p^4$-Lagrangian involves ten free
parameters which were determined in the fundamental papers of Gasser and
Leutwyler \cite{GasserLeutw1}. For the $p^6$ Lagrangian there are
already more than a hundred of them \cite{FearingScherer}. Many
interesting low energy hadronic amplitudes have now been calculated to
order $p^4$ and to order $p^{6}$ \cite{neu} in the case of \ 
$SU(2)\times SU(2)$ \ chiral perturbation theory which involves only one
mass scale. Recent progress in the calculation of massive two-loop
integrals allows now calculations to order $p^{6}$ in the full
$SU(3)\times SU(3)$ \ chiral perturbation theory. With so many unknown
parameters, however, one may ask about the usefulness of such
calculations. Here are a few arguments in favour:
\begin{itemize}
\item In a given class of experiments, such as the electromagnetic and
  weak form factors of the light mesons, only a limited number of
  renormalization constants enter and relations between amplitudes can
  be tested \cite{PostSchilcher}.
\item The unknown constants enter only polynomially and precision
  experiments could separate the unambiguous predictions (see e.g. the
  subsequent discussion).
\item Knowledge of the exact low-energy functional form of an
  amplitude may be important for the experimental extraction of
  low-energy parameters such as charge radii.
\item The results may be used in model calculations which predict the
  polynomial terms. These calculations can then be compared with
  experiment.
\item The question of convergence of the chiral perturbation theory
  may be addressed.
\end{itemize}

We have embarked on the program of a full $p^6$ and two-loop analysis
of the various semi-leptonic form factors of pions and kaons in full
$SU(3)\times SU(3)$ chiral perturbation theory. From the point of view
of practicality it seems prudent to begin such an ambitious program
with the simplest process possible involving the fewest number of
renormalization parameters. Here this would be the electromagnetic
form factor of the neutral kaon. There exists no tree level
contribution to ${\cal O}(p^4)$ and therefore the one-loop
contributions must be finite. At ${\cal O}(p^6)$, however, a tree
level term exists which couples the $K^0$ directly to the
electromagnetic field tensor of the external photon, and a new
renormalization parameter enters. In this paper we will present the results
of our ${\cal O}(p^6)$ calculation of the neutral kaon's
electromagnetic form factor. It is defined by the matrix element of
the electromagnetic current between neutral kaon states:
\begin{eqnarray}
  \label{DefCurrentMatrixElem}
  \langle K^0,p^\prime \mid J_\mu \mid K^0,p \rangle
  &=& (p+p^\prime)_\mu \; F^{K^0}(q^2) \:,
\end{eqnarray}
where $q=p\f{-}p^\prime$ is the momentum of the photon  and $J_\mu$ stands for
the electromagnetic current carried by the light quarks which is a linear
combination of the chiral currents $V_\mu^3$ and $V_\mu^8$:
\begin{equation}
  J_\mu
  \;\:\: = \;\:\:
  \frac{2}{3} \bar{u}\gamma_\mu u
  - \frac{1}{3} \bar{d}\gamma_\mu d
  - \frac{1}{3} \bar{s}\gamma_\mu s
  \;\:\: = \;\:\:
  V_\mu^3 + \frac{1}{\sqrt{3}} \, V_\mu^8 \:.
\end{equation}
We neglect the contributions of the heavy quarks.

%%%%%%%%%%%%%%%%%%%%%%%%%%%%%%%%%%%%%%%%%%%%%%%%%%%%%%%%%%%%%%%%%%%%%%%%%%%
%%%%%%%%%%%%%%%%%%%%%%%%%%%%%%%%%%%%%%%%%%%%%%%%%%%%%%%%%%%%%%%%%%%%%%%%%%%

\section{The effective Lagrangian and the diagrams}

Chiral perturbation theory is formulated in terms of an effective
Lagrangian involving an increasing number of covariant derivatives as
well as external sources,
\begin{eqnarray}
  {\cal L}_{\mbox{\scriptsize\em eff}} &=&
  {\cal L}^{(2)} + {\cal L}^{(4)} + {\cal L}^{(6)} + \ldots
\end{eqnarray}
where
\begin{eqnarray}
  {\cal L}^{(2)} &=&
  \hphantom{+}
  \frac{F^2}{4} \: \mbox{\rm Tr}\big( D_\mu\!\!\;U D^\mu\!\!\:U^\dagger \big)
  + \frac{F^2}{4} \: \mbox{\rm Tr}\big( \chi U^\dagger + U \chi^\dagger \big)
  \\
  {\cal L}^{(4)} &=&
  L_1 \: \{ \mbox{\rm Tr}\big( D_\mu\!\!\;U D^\mu\!\!\:U^\dagger \big) \}^2
  + L_2 \: \mbox{\rm Tr}\big( D_\mu\!\!\;U D_\nu\!\!\;U^\dagger \big) \;
  \mbox{\rm Tr}\big( D^\mu\!\!\:U D^\nu\!\!\:U^\dagger \big)
  \\&&
  + \; L_3 \: \mbox{\rm Tr}\big( D_\mu\!\!\;U D^\mu\!\!\:U^\dagger
  D_\nu\!\!\;U D^\nu\!\!\:U^\dagger \big)
  + L_4 \: \mbox{\rm Tr}\big( D_\mu\!\!\;U D^\mu\!\!\:U^\dagger \big) \;
  \mbox{\rm Tr}\big( \chi U^\dagger + U \chi^\dagger\big)
  \nonumber\\&&
  + \; L_5 \: \mbox{\rm Tr}\big( D_\mu\!\!\;U D^\mu\!\!\:U^\dagger
  \big( \chi U^\dagger + U \chi^\dagger\big)\big)
  + L_6 \: \big\{
  \mbox{\rm Tr}\big( \chi U^\dagger + U \chi^\dagger\big) \big\}^2
  \nonumber\\&&
  + \; L_7 \: \big\{
  \mbox{\rm Tr}\big( \chi^\dagger U - U^\dagger \chi\big) \big\}^2
  + L_8 \: \mbox{\rm Tr}\big( \chi U^\dagger \chi U^\dagger
  + U \chi^\dagger U \chi^\dagger \big)
  \nonumber\\&&
  - \; i L_9 \: \mbox{\rm Tr}\big( L_{\mu\nu} D^\mu\!\!\:U D^\nu\!\!\:U^\dagger
  + R_{\mu\nu} D^\mu\!\!\:U^\dagger D^\nu\!\!\:U \big)
  + L_{10} \: \mbox{\rm Tr}\big( L_{\mu\nu} U R^{\mu\nu} U^\dagger \big)
  \nonumber
  \\
  \label{L6LagrangianRel}
  {\cal L}^{(6)} &=& \frac{\beta}{F^2} \, F_{\mu\nu} \,
  \mbox{\rm Tr}\big(\partial_\mu\!\!\;U \partial^\mu\!\!\:U^\dagger \big)
  + \ldots \:,
\end{eqnarray}
$U(x)=\exp[i\phi(x)/F]$ is the unitary $3\times 3$ matrix made up of
the Goldstone fields, $F$ is the pion decay constant in the chiral
limit, $D_\mu$ the covariant derivative involving external vector- and
axial vector sources of the corresponding currents, $F_{\mu\nu}$ is the
electromagnetic field tensor, and $\chi$ is related to the quark mass matrix.
In ${\cal L}^{(6)}$ we have only displayed the terms relevant to the $K^0$ form
factor. The ${\cal L}^{(6)}$ term is extracted from reference
\cite{FearingScherer} where in their notation
$\beta = \frac{4}{3} F^2 (m_K^2 - m_\pi^2) ( 2 B_{24} - B_{25})$.

Each term in ${\cal L}_{\mbox{\scriptsize\em eff}}$ produces
vertices of $2,4,6,\ldots$ mesons with or without attached photon.
According to Weinberg's power counting theorem \cite{Weinberg}
the form factor at order $p^6$ of the chiral expansion is the sum
of all 2-loop diagrams with vertices from ${\cal L}^{(2)}$,
all 1-loop diagrams with vertices from ${\cal L}^{(2)}$ and
one vertex from ${\cal L}^{(4)}$, and a tree graph with two vertices from
${\cal L}^{(4)}$ or one vertex from ${\cal L}^{(6)}$. 
Each diagram has two external meson lines
representing the incoming and outgoing $K^0$ and one external photon
line. All relevant diagrams for the unrenormalized $K^0$ form factor
are given in fig.~1. To indicate the origin of the
vertices in the Feynman diagrams we use the following notation:
${\cal L}^{(2)}$-vertices are denoted by filled circles
    \parbox{1.2ex}{ \unitlength1ex\begin{picture}(0,0)
                  \put(0.5,0){\circle*{1}}\end{picture} },
${\cal L}^{(4)}$-vertices by filled squares
    \parbox{1ex}{ \unitlength1ex\begin{picture}(0,0)
                \put(0.5,0){\makebox(0,0){\rule[0ex]{1ex}{1ex}}}
                \end{picture} }
and an ${\cal L}^{(6)}$-vertex by an open square
    \parbox{1.2ex}{ \unitlength1ex\begin{picture}(0,0)
                  \put(0.5,0){\begin{picture}(0,0)\unitlength0.5ex
                    \put(-1,1){\line(1,0){2}}\put(-1,1){\line(0,-1){2}}
                    \put(1,1){\line(0,-1){2}}\put(-1,-1){\line(1,0){2}}
                  \end{picture}}\end{picture} }.
Only the ${\cal L}^{(6)}$ vertex is new.

%~~~~~~~~~~~~~~~~
%  FIGURE 1
%~~~~~~~~~~~~~~~~

\begin{figure}
  \centerline{
    \includegraphics*[scale=0.8]{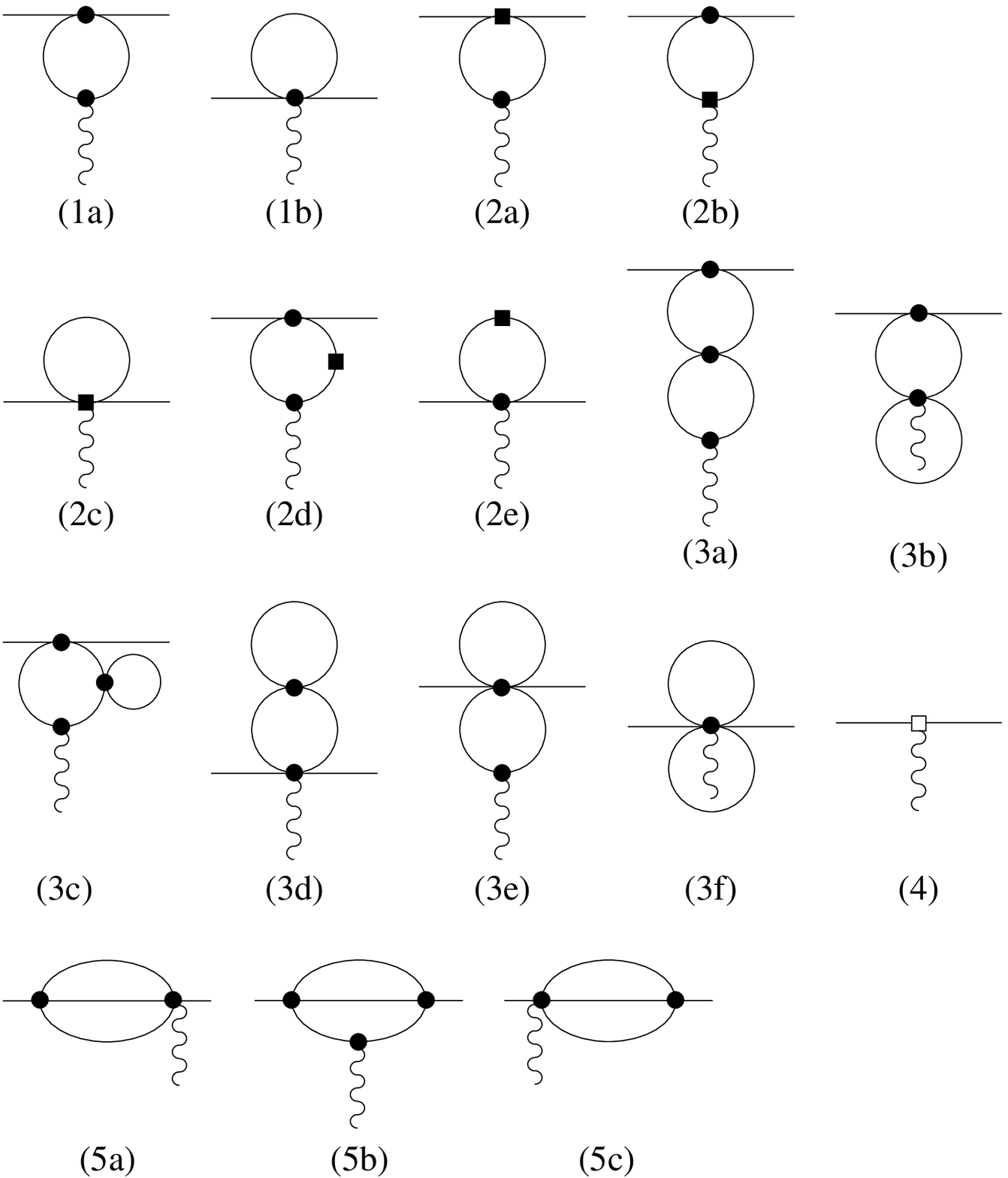}
  }
  Figure 1:
  \footnotesize
  The diagrams for the $K^0$-form factor up to order $p^6$.
   ${\cal L}^{(2)}$-vertices are denoted by filled circles
     (\parbox{1.2ex}{ \unitlength1ex\begin{picture}(0,0)
                   \put(0.5,0){\circle*{1}}\end{picture} }),
  ${\cal L}^{(4)}$-vertices by filled squares
    (\parbox{1ex}{ \unitlength1ex\begin{picture}(0,0)
                   \put(0.5,0){\makebox(0,0){\rule[0ex]{1ex}{1ex}}}
                   \end{picture} }),
  and an ${\cal L}^{(6)}$-vertex by an open square
    (\parbox{1.2ex}{ \unitlength1ex\begin{picture}(0,0)
                     \put(0.5,0){\begin{picture}(0,0)\unitlength0.5ex
                       \put(-1,1){\line(1,0){2}}\put(-1,1){\line(0,-1){2}}
                       \put(1,1){\line(0,-1){2}}\put(-1,-1){\line(1,0){2}}
                     \end{picture}}\end{picture} }).
\end{figure}

In each loop diagram, the internal lines symbolize an arbitrary
$\pi$, $K$, or $\eta$ meson and they must be summed over.
Due to the derivative couplings of ${\cal L}_{\mbox{\scriptsize\em eff}}$
the Feynman rules of the vertices are quite involved.
For the relevant vertices they are discussed in appendix \ref{FeynmanRules}.

The diagrams (1a) and (1b) represent the well-known result of order $p^4$
which is the leading order approximation of the $K^0$ form factor
\cite{GasserLeutw1}. The pure $p^6$ diagrams consist of two groups:
\begin{itemize}
  \item the reducible diagrams, i.e.
       the 1-loop diagrams (2a)--(2e), the 2-loop diagrams (3a)--(3f),
       where the loops are independent of each other,
       and the tree graph (4)
  \item the irreducible 2-loop diagrams (5a)--(5c).
\end{itemize}
All diagrams of the first group can be reduced to elementary 1-loop
integrals. The results are presented in section \ref{ReducibleDiagrams}.
The calculation of the irreducible 2-loop diagrams is much harder
to carry out: since in $SU(3)$ chiral perturbation theory there are three
different mass scales of the same order of magnitude ($m_\pi$, $m_K$,
$m_\eta$), these 2-loop integrals can no longer be expressed by elementary
analytical functions. In section \ref{IrreducibleDiagrams}
we discuss their contributions to the $K^0$ form factor, and in 
appendix \ref{CalcOfIrrInt} we describe our numerical algorithm for
the evaluation of these diagrams.

The contributions of the diagrams can be classified further according to their 
dependence on the momentum $q$ of the photon. Diagrams, where the
photon momentum does not flow through, are polynomials in $q^2$:
(1b), (2e), (3d), (5a), and (5c) do not depend on $q^2$, (2c) is a polynomial 
in $q^2$ of degree $\leq 1$, and (4) has at most degree 2.
Since the normalization of the $K^0$ form factor is fixed by the charge
of the particle due to the Ward-Fradkin-Takahashi identity 
\cite{WardFradkinTakahashi},
\bea
  F^{K^0}(0) &=& 0 \:,
\eea
the interesting contributions are those with genuine $q^2$ dependence.
Nevertheless, we calculate all diagrams, because this procedure offers
additional possibilities to check the calculation.

%%%%%%%%%%%%%%%%%%%%%%%%%%%%%%%%%%%%%%%%%%%%%%%%%%%%%%%%%%%%%%%%%%%%%%%%%%%
%%%%%%%%%%%%%%%%%%%%%%%%%%%%%%%%%%%%%%%%%%%%%%%%%%%%%%%%%%%%%%%%%%%%%%%%%%%

\section{Reducible diagrams}
\label{ReducibleDiagrams}

The diagrams (1a)--(3f) are 1-loop integrals or products thereof
and can be calculated analytically. Inserting the Feynman rules
into the vertices yields integrands of the structure
\bea\label{structureOfIntegrands}
  \frac{V(k^2,kp_1,kp_2,kl,l^2,lp_1,lp_2)}{P}
\eea
where $p_1$ and $p_2$ are the external momenta of the incoming and the outgoing
$K^0$, $k$ and $l$ are the internal loop momenta, the denominator $P$
is a product of scalar propagator factors depending on the topology
of the diagram, and the numerator $V$ is a polynomial coming from the
vertex factors. These integrals can be expressed by a set of basic 1-loop
functions in the following way:
First, $p_1$ and $p_2$ are transformed into $p \f{=} p_1 \f{+} p_2$
and the momentum $q \f{=} p_1 \f{-} p_2$ of the photon.
Then, some factors in the numerator are canceled against some
propagator factors in the denominator via
\bea\label{erste_Umformung}
k^2 &=& P(k,m^2)+m^2
\\
kq &=& \frac{1}{2} \Big[ P(k\f{+}q,m^2)+m^2-k^2-q^2 \Big]
\\
l^2 &=& P(l,m^2)+m^2
\\
lq &=& \frac{1}{2} \Big[ P(l\f{+}q,m^2)+m^2-l^2-q^2 \Big]
\eea
with $P(k,m) \f{=} k^2 \f{-} m^2$.
What remains are 1-loop 1-point and 2-point functions with tensor
numerators which are decomposable into Lorentz covariants.
As our calculation shows, there are in fact only two 1-loop functions
by means of which all reducible loop contributions of the $K^0$ form factor
can be expressed: On the one hand the scalar 1-point function with
its derivative,
\begin{equation}
  A(m^2)
  \;\:=\;\:
  \parbox{12mm}{
    \unitlength1mm
    \begin{picture}(12,18)
    \put(6,9){\begin{picture}(0,0)\thicklines\unitlength1mm
       \put(0,0){\circle{10}}
       \put(0,-5){\makebox(0,0){\scriptsize$\bullet$}}
       %\put(0,6){\makebox(0,0)[b]{$k$,$m$}}
       \put(0,6){\makebox(0,0)[b]{$m$}}
       \end{picture}}
    \end{picture}
  }
  \quad\;\mbox{and}\quad\;
  A^\prime(m^2)
  \;\:=\;\:
  \frac{\partial A}{\partial m^2} \:,
\end{equation}
and on the other hand one specific tensor coefficient of the 2-point function,
also with derivative,
\begin{equation}
  B(q^2,m^2) \;\:=\;\:
  \parbox{28mm}{
    \unitlength1mm
    \begin{picture}(28,18)
    \put(14,9){\begin{picture}(0,0)\thicklines\unitlength1mm
       \put(0,0){\circle{10}}
       \put(-5,0){\line(-1,0){8}}
       \put(5,0){\line(1,0){8}}
       \put(-5,0){\makebox(0,0){\scriptsize$\bullet$}}
       \put(5,0){\makebox(0,0){\scriptsize$\bullet$}}
       \put(0,6){\makebox(0,0)[b]{$m$}}
       \put(0,-6){\makebox(0,0)[t]{$m$}}
       \put(-11,1){\makebox(0,0)[b]{$q$}}
       \end{picture}}
    \end{picture}
  }
  \quad\;\mbox{and}\quad\;
  B^\prime(q^2,m^2)
  \;\:=\;\:
  \frac{1}{2} \frac{\partial B}{\partial m^2} \:.
\end{equation}
These basic 1-loop integrals are defined in detail in appendix
\ref{relevantOneLoopIntegrals}.
In the following equations we give the contribution $\Delta$
of each reducible diagram to the $K^0$
form factor in terms of the basic functions $A$ and $B$:
\begin{eqnarray}
  %~~~~~~~~~~~~~~~~
  %  diagram (1a)
  %~~~~~~~~~~~~~~~~
  \label{Einzelbeitraege_der_reduziblen_diagrame}
  \Diag{1a}{}{K^0}{} &=& \frac{1}{\vierpi{2} F^2} \;
  \Big\{ \BOne{2}{1}{q^2}{m_\pi^2}{m_\pi^2} - \BOne{2}{1}{q^2}{m_\Ka^2}{m_\Ka^2} \Big\}
\\
  %~~~~~~~~~~~~~~~~
  %  diagram (1b)
  %~~~~~~~~~~~~~~~~
  \Diag{1b}{}{K^0}{} &=&
  -\frac{1}{2\vierpi{2} F^2} \; \Big\{ A(m_\pi^2) - A(m_\Ka^2) \Big\}
\\
  %~~~~~~~~~~~~~~~~
  %  diagram (2a)
  %~~~~~~~~~~~~~~~~
  \Diag{2a}{}{K^0}{} &=&
  \frac{4}{\vierpi{2} F^4} \; \Big\{ \BOne{2}{1}{q^2}{m_\pi^2}{m_\pi^2}
  \big[ 2(m_\pi^2+2m_\Ka^2) L_4 + 2(m_\pi^2+m_\Ka^2) L_5 - q^2 L_3 \big]
  \\
  &&\qquad - \BOne{2}{1}{q^2}{m_\Ka^2}{m_\Ka^2}
  \big[ 2(m_\pi^2+2m_\Ka^2) L_4 + 4m_\Ka^2 L_5 - q^2 L_3 \big] \Big\}
  \nn
\\
  %~~~~~~~~~~~~~~~~
  %  diagram (2b)
  %~~~~~~~~~~~~~~~~
  \Diag{2b}{}{K^0}{} &=&
  \frac{2}{\vierpi{2} F^4} \; \Big\{ \BOne{2}{1}{q^2}{m_\pi^2}{m_\pi^2}
  \big[ 4(m_\pi^2+2m_\Ka^2) L_4 + 4m_\pi^2 L_5 + q^2 L_9 \big]
  \\
  &&\qquad  - \BOne{2}{1}{q^2}{m_\Ka^2}{m_\Ka^2}
  \big[ 4(m_\pi^2+2m_\Ka^2) L_4 + 4m_\Ka^2 L_5 + q^2 L_9 \big]
  \Big\}
  \nn
\\
  %~~~~~~~~~~~~~~~~
  %  diagram (2c)
  %~~~~~~~~~~~~~~~~
  \Diag{2c}{}{K^0}{} &=&
  -\frac{1}{\vierpi{2} F^4} \; \Big\{ A(m_\pi^2)
  \big[ 4(m_\pi^2+2m_\Ka^2) L_4 + 4(m_\pi^2+m_\Ka^2) L_5 + q^2 L_9 \big]
  \\
  &&\qquad  - A(m_\Ka^2)
  \big[ 4(m_\pi^2+2m_\Ka^2) L_4 + 8 m_\Ka^2 L_5 + q^2 L_9 \big]
  \Big\}
  \nn
\\
  %~~~~~~~~~~~~~~~~
  %  diagram (2d)
  %~~~~~~~~~~~~~~~~
  \Diag{2d}{}{K^0}{} &=&
  \frac{16}{F^4} \; \Big\{
   \BOne{2}{1}{q^2}{m_\Ka^2}{m_\Ka^2}
  \: [L_4 (2 m_\Ka^2 + m_\pi^2) + L_5 m_\Ka^2 ]
  \\
  &&\qquad
  - \BOne{2}{1}{q^2}{m_\pi^2}{m_\pi^2}
  \: [ L_4 (2 m_\Ka^2 + m_\pi^2) + L_5 m_\pi^2]
  \nn\\
  &&\hspace{-0em}
  + \BOneNenner{2}{1}{q^2}{1}{m_\Ka^2}{2}{m_\Ka^2}
  \: m_\Ka^2 \:[
  L_4 ( 2 m_\Ka^2 + m_\pi^2)
  +L_5 m_\Ka^2
  -2 L_6 (2 m_\Ka^2 + m_\pi^2)
  -2 L_8 m_\Ka^2]
  \nn\\
  &&\hspace{-0em}
  - \BOneNenner{2}{1}{q^2}{1}{m_\pi^2}{2}{m_\pi^2}
  \: m_\pi^2 \:[
   L_4 (2 m_\Ka^2 + m_\pi^2)
  +L_5 m_\pi^2
  -2 L_6 (2 m_\Ka^2 + m_\pi^2)
  -2 L_8 m_\pi^2]
  \Big\}\nn
\\
  %~~~~~~~~~~~~~~~~
  %  diagram (2e)
  %~~~~~~~~~~~~~~~~
  \Diag{2e}{}{K^0}{} &=&
  \frac{1}{F^4}  \Big\{
  4 A(m_\pi^2) \: [L_4 (2 m_\Ka^2+m_\pi^2)+L_5 m_\pi^2]
  \\
  &&\qquad
  -4 A(m_\Ka^2) \: [L_4 (2 m_\Ka^2 + m_\pi^2) + L_5 m_\Ka^2 ]
  \nn\\
  &&\qquad
  -4 \ANenner{2}{m_\Ka^2} \: m_\Ka^2 \:[
   L_4 (2 m_\Ka^2 + m_\pi^2)
  +L_5 m_\Ka^2
  -2 L_6 (2 m_\Ka^2 + m_\pi^2)
  -2 L_8 m_\Ka^2]
  \nn\\
  &&\qquad
  +4 \ANenner{2}{m_\pi^2} \: m_\pi^2 \:[
  L_4 (2 m_\Ka^2 + m_\pi^2)
  +L_5 m_\pi^2
  -2 L_6 (2 m_\Ka^2 + m_\pi^2)
  -2 L_8 m_\pi^2]
  \Big\}\nn
\\
  %~~~~~~~~~~~~~~~~
  %  diagram (3a)
  %~~~~~~~~~~~~~~~~
  \Diag{3a}{}{K^0}{} &=&
  -\frac{2}{\vierpi{4} F^4} \; \Big\{
  \BOne{2}{1}{q^2}{m_\pi^2}{m_\pi^2}^2 + \BOne{2}{1}{q^2}{m_\pi^2}{m_\pi^2} \: \BOne{2}{1}{q^2}{m_\Ka^2}{m_\Ka^2}
  - 2 \BOne{2}{1}{q^2}{m_\Ka^2}{m_\Ka^2}^2 \Big\}
  \nn
\\
  %~~~~~~~~~~~~~~~~
  %  diagram (3b)
  %~~~~~~~~~~~~~~~~
  \Diag{3b}{}{K^0}{} &=&
  \frac{1}{12\vierpi{4} F^4} \; \Big\{
  20 \: A(m_\pi^2) \: \BOne{2}{1}{q^2}{m_\pi^2}{m_\pi^2}
  + 10 \: A(m_\Ka^2) \: \BOne{2}{1}{q^2}{m_\pi^2}{m_\pi^2}
  \\
  &&\qquad  + 3 \: A(m_\pi^2) \: \BOne{2}{1}{q^2}{m_\Ka^2}{m_\Ka^2}
  - 30 A(m_\Ka^2) \: \BOne{2}{1}{q^2}{m_\Ka^2}{m_\Ka^2}
  \nonumber\\
  &&\qquad - 3 \: A(m_\eta^2) \: \BOne{2}{1}{q^2}{m_\Ka^2}{m_\Ka^2} \Big\}
  \nn
\\
  %~~~~~~~~~~~~~~~~
  %  diagram (3c)
  %~~~~~~~~~~~~~~~~
  \Diag{3c}{}{K^0}{} &=&
  \frac{1}{6 F^4}  \; \Big\{
  3 \: \BOne{2}{1}{q^2}{m_\Ka^2}{m_\Ka^2}
  \: \big[ A(m_\eta^2) + 2 \: A(m_\Ka^2) + A(m_\pi^2) \big]
  \\
  &&\qquad
  -4 \: \BOne{2}{1}{q^2}{m_\pi^2}{m_\pi^2} \:
  \big[ A(m_\Ka^2) + 2 \: A(m_\pi^2) \big]
  \nn\\
  &&\qquad
  +2 \: \BOneNenner{2}{1}{q^2}{1}{m_\pi^2}{2}{m_\pi^2} \:
  m_\pi^2 \big[  A(m_\eta^2) - 3\, A(m_\pi^2) \big]
  \nn\\
  &&\qquad
  +\BOneNenner{2}{1}{q^2}{1}{m_\Ka^2}{2}{m_\Ka^2}
  \: A(m_\eta^2) \: (3 m_\eta^2+m_\pi^2)
  \Big\}\nn
\\
  %~~~~~~~~~~~~~~~~
  %  diagram (3d)
  %~~~~~~~~~~~~~~~~
  \Diag{3d}{}{K^0}{} &=&
  -\frac{1}{24 F^4}  \; \Big\{
  3 \: A(m_\eta^2) \: A(m_\Ka^2)
  +6 \: A(m_\Ka^2)^2
  -A(m_\Ka^2) \: A(m_\pi^2)
  -8 \: A(m_\pi^2)^2
  \\
  &&\qquad
  +2 m_\pi^2 \: \ANenner{2}{m_\pi^2} \: A(m_\eta^2)
  -6 m_\pi^2\: \ANenner{2}{m_\pi^2} \: A(m_\pi^2)
  \nn\\
  &&\qquad
  + (3 m_\eta^2+m_\pi^2) \: \ANenner{2}{m_\Ka^2} \: A(m_\eta^2)
  \Big\}\nn
\\
  %~~~~~~~~~~~~~~~~
  %  diagram (3e)
  %~~~~~~~~~~~~~~~~
  \Diag{3e}{}{K^0}{} &=&
  \frac{1}{6\vierpi{4} F^4} \; \Big\{
  5 A(m_\pi^2) \: \BOne{2}{1}{q^2}{m_\pi^2}{m_\pi^2}
  + 4 A(m_\Ka^2) \: \BOne{2}{1}{q^2}{m_\pi^2}{m_\pi^2}
  \\
  &&\qquad  + A(m_\eta^2) \: \BOne{2}{1}{q^2}{m_\pi^2}{m_\pi^2}
  - A(m_\pi^2) \: \BOne{2}{1}{q^2}{m_\Ka^2}{m_\Ka^2}
  \nonumber\\
  &&\qquad
  - 8 A(m_\Ka^2) \: \BOne{2}{1}{q^2}{m_\Ka^2}{m_\Ka^2}
  - A(m_\eta^2) \: \BOne{2}{1}{q^2}{m_\Ka^2}{m_\Ka^2} \Big\}
  \nn
\\
  %~~~~~~~~~~~~~~~~
  %  diagram (3f)
  %~~~~~~~~~~~~~~~~
  \Diag{3f}{}{K^0}{} &=&
  -\frac{1}{12\vierpi{4} F^4} \; \Big\{
  5 A(m_\pi^2)^2
  + 3 A(m_\Ka^2) \: A(m_\pi^2)
  + A(m_\eta^2) \: A(m_\pi^2)
  \\
  &&\qquad
  - 8 A(m_\Ka^2)^2
  - A(m_\eta^2) \: A(m_\Ka^2) \Big\}
  \nn
\\
  %~~~~~~~~~~~~~~~
  %  diagram (4)
  %~~~~~~~~~~~~~~~
  \label{diag4b}
  \Diag{4}{}{K^0}{} &=& \frac{\beta q^2}{F^4} \:.
\end{eqnarray}
The contributions contain three masses $m_\pi$, $m_K$, and $m_\eta$,
but in the isospin limit only two of them are independent.
$m_\eta^2$ is a short-hand notation for the Gell-Mann-Okubo
relation $\frac{4}{3}m_K^2\f{-}\frac{1}{3}m_\pi^2$ 
which corresponds to the mass of the $\eta$ meson in leading order 
${\cal O}(p^2)$.

Since at order ${\cal O}(p^2)$ the $K^0$ form factor vanishes,
the only diagrams which are subject to renormalization are the
${\cal O}(p^4)$ diagrams (1a) and (1b). Due to renormalization there
are three modifications of these diagrams:
\begin{itemize}
  \item Wave function renormalization of the outer $K^0$ lines
        requires an additional factor $1 + \delta Z_K$
        where $\delta Z_K$ is the well-known ${\cal O}(p^4)$ result
        (cf.\ \cite{GasserLeutw1})
        \begin{eqnarray}
          \delta Z_\Ka &=&
          -\frac{1}{4F^2} \Big\{
          A(m_\eta^2)+2 A(m_\Ka^2)+A(m_\pi^2)
          + 32 L_4 (2 m_\Ka^2 + m_\pi^2) + 32 L_5 m_\Ka^2 \Big\}
          \:.\qquad
          \label{ZKa_zu_1loop}
        \end{eqnarray}
  \item Renormalization of the pion decay constant $F_\pi$ has to be
        taken into account: the plain ${\cal O}(p^4)$ result contains
        a factor $1/F^2$ where $F$ can no longer be identified
        with $F_\pi$ at order $p^6$. Instead, one has to consider
        the ${\cal O}(p^4)$ correction
        $F_\pi \f{=} F ( 1 \f{+} \delta\!\!\:f)$ with (cf. \cite{GasserLeutw1})
        \bea
          \delta\!\!\: f &=&
          \frac{1}{2F^2} \Big\{
          A(m_\Ka^2) + 2 A(m_\pi^2)
          + 8 L_4 (2 m_\Ka^2 + m_\pi^2) + 8 L_5 m_\pi^2 \Big\} \:.
        \eea
  \item Analogously, the mass renormalizations are to be included, i.e. the
        masses $m$ appearing in
        (\ref{Einzelbeitraege_der_reduziblen_diagrame}f\/f)
        are related to the physical masses $m_{\mbox{\scriptsize\em ph}}$ via
        $m^2_{\mbox{\scriptsize\em ph}} \f{=} m^2 \f{+}
        \Sigma(m^2_{\mbox{\scriptsize\em ph}})$
        where $\Sigma$ stands for the ${\cal O}(p^4)$ self energies
        (cf. \cite{GasserLeutw1})
        \bea
          -\Sigma_\pi(m_\pi^2) &=&
          \frac{1}{6 F^2} \: \Big\{
          -A(m_\eta^2) \: m_\pi^2
          +3 A(m_\pi^2) \: m_\pi^2
          +48 L_4 m_\pi^2 (2 m_\Ka^2+m_\pi^2)
          \\&&\qquad
          +48 L_5 m_\pi^4
          +96 L_6 m_\pi^2 (-2 m_\Ka^2-m_\pi^2)
          -96 L_8 m_\pi^4
          \Big\}\nn
          \\
          -\Sigma_\Ka(m_\Ka^2) &=&
          \frac{1}{12 F^2} \: \Big\{
          A(m_\eta^2) \: (3 m_\eta^2+m_\pi^2)
          +96 L_4 m_\Ka^2 (2 m_\Ka^2+m_\pi^2)
          +96 L_5 m_\Ka^4
          \\&&\qquad
          +192 L_6 m_\Ka^2 (-2 m_\Ka^2-m_\pi^2)
          -192 L_8 m_\Ka^4
          \Big\} \:. \nn
        \eea
\end{itemize}
In summary, the total contribution of diagrams (1a) and (1b) to the
$K^0$ form factor at order $p^6$ is given by
\bea
 \Big( 1 + \delta Z_K \Big)
 \Big( \Diag{1a}{}{K^0}{} + \Diag{1b}{}{K^0}{} \Big)
 \Big|_{ m^2 = m_{\mbox{\tiny\em ph}}^2-\Sigma(m_{\mbox{\tiny\em ph}}^2),\;
 F=F_\pi(1-\delta\!\!\: f)} \:.
 \label{RenormalizationCorr}
\eea
Renormalization corrections of the other diagrams are of order
$p^8$ and can be neglected. 

The sum of all reducible diagrams including the above renormalization
corrections takes the following compact form:
\newcommand{\AInt}[1]{A(m_{#1}^2)}
\newcommand{\BInt}[1]{B(q^2,m_{#1}^2)}
\bea
  F^4 \Delta^{\mbox{\scriptsize\em (red)}}
  &=&
  \frac{1}{24} \Big\{
  - \AInt{\eta} \: \Big[ \, 4 \: \AInt{\Ka} - 8 \: \BInt{\Ka} 
  - \AInt{\pi} + 2 \: \BInt{\pi} \Big]
  \\
  &&\qquad\quad
  + \; 16 \: \AInt{\Ka}^2 
  + 10 \: \AInt{\Ka} \: \AInt{\pi} 
  - 23 \: \AInt{\pi}^2
  - 80 \: \AInt{\Ka} \: \BInt{\Ka} 
  \nn\\
  &&\qquad\quad
  + \; 32 \: \AInt{\Ka} \: \BInt{\pi} 
  - 28 \: \AInt{\pi} \: \BInt{\Ka} 
  + 70 \: \AInt{\pi} \: \BInt{\pi}
  \nn\\
  &&\qquad\quad
  + \; 96 \: \BInt{\Ka}^2 
  - 48 \: \BInt{\Ka} \: \BInt{\pi} 
  - 48 \: \BInt{\pi}^2
  \Big\}
  \nn\\
  &&
  + \; 4 \: q^2 \: L_3 \: \Big[ \BInt{\Ka} - \BInt{\pi} \Big]
  \nn\\
  &&
  + \; ( 4 \: m_\pi^2 \: L_5 + q^2 \: L_9 )
    \: \Big[ \AInt{\Ka} - 2 \: \BInt{\Ka} - \AInt{\pi} + 2 \: \BInt{\pi}
    \Big] \:.
  \nn
\eea
Note that the mass derivatives $A^\prime$ and $B^\prime$ 
and the ${\cal L}^{(4)}$ parameters $L_4$, $L_6$, and $L_8$
which occurred in some diagrams have vanished. The only 
${\cal L}^{(4)}$ parameters which are relevant for the $K^0$ form factor
at order $p^6$ are $L_3$, $L_5$, and $L_9$.

%%%%%%%%%%%%%%%%%%%%%%%%%%%%%%%%%%%%%%%%%%%%%%%%%%%%%%%%%%%%%%%%%%%%%%%%%%%
%%%%%%%%%%%%%%%%%%%%%%%%%%%%%%%%%%%%%%%%%%%%%%%%%%%%%%%%%%%%%%%%%%%%%%%%%%%

\section{Irreducible diagrams}
\label{IrreducibleDiagrams}
In the irreducible diagrams (5a)--(5c) the 2-loop integrations
are not independent of each other as they were in the reducible graphs.
That is why genuine 2-loop functions enter the stage which cannot be expressed
by 1-loop integrals only.

Inserting the Feynman rules yields integrands with a similar structure as in
(\ref{structureOfIntegrands}), e.g. for diagram (5b)
\begin{equation}
  \frac{V(k^2,kp_1,kp_2,kl,l^2,lp_1,lp_2)}
  {P(k\f{+} p_1,m_1^2) \: P(k \f{+} p_2,m_1^2) \:
  P(k \f{+} l,m_2^2) \: P(l,m_3^2)}
  \:,
\end{equation}
where $V$ is a polynomial of degree equal to the number of vertices.
After canceling factors via
\bea
  kp_1 &=& \frac{1}{2} \Big[ P(k\f{+}p_1,m_1^2)+m_0^2-k^2-p_1^2 \Big]
  \\
  kp_2 &=& \frac{1}{2} \Big[ P(k\f{+}p_2,m_1^2)+m_1^2-k^2-p_2^2 \Big]
  \\
  kl &=& \frac{1}{2} \Big[ P(k\f{+}l,m_2^2)+m_2^2-k^2-l^2 \Big]
  \\
  l^2 &=& P(l,m_3^2)+m_3^2
\eea
we are left with reducible integrals which can be calculated analytically,
and with some genuine 2-loop integrals of the \emph{sunset}-topology,
i.e. the 3-point functions
\bea
  \label{Def_von_TFull}
  T_{\alpha_1, \alpha_2, \beta}(q^2;p_1^2,p_2^2;m_0^2,m_1^2,m_2^2,m_3^2)
  &=&
  \nonumber\\
  &&\hspace{-13em} \mu^{8-2D} \int \frac{d^D\!\!\:k}{i(2\pi)^D}
  \; \frac{d^D\!\!\:l}{i(2\pi)^D} \; \frac{
  (lp_1)^{\alpha_1} \; (lp_2)^{\alpha_2} \; (k^2)^\beta}{
  P(k\f{+}p_1,m_0^2) \; P(k\f{+}p_2,m_1^2) \; P(k\f{+}l,m_2^2) \; P(l,m_3^2)}
  \quad\hspace{3em}
\eea
and the 2-point functions
\bea
  S_{\alpha, \beta}(p^2; m_1^2,m_2^2,m_3^2)
  &=&
  \mu^{8-2D} \int \frac{d^D\!\!\:k}{i(2\pi)^D} \;
  \frac{d^D\!\!\:l}{i(2\pi)^D} \; \frac{ (lp)^{\alpha} \; (k^2)^\beta}{
  P(k\f{+}p,m_1^2) \; P(k\f{+}l,m_2^2) \; P(l,m_3^2)}
  \:. \hspace{3em}
\eea

In diagram (5b), five different mass flows of intermediate mesons
must be regarded:
\be
  \Diag{5b}{}{K^0}{} \;\:=\;\: \Diag{5b}{\Ka\pi\pi}{K^0}{}
                             + \Diag{5b}{\Ka\pi\eta}{K^0}{}
                             + \Diag{5b}{\Ka\Ka\Ka}{K^0}{}
                             + \Diag{5b}{\pi \Ka\pi}{K^0}{}
                             + \Diag{5b}{\pi \Ka\eta}{K^0}{}
                             \:, \quad
\ee
where $\Diag{5b}{rst}{K^0}{}$ means that meson $r$ couples to the
photon and the other two lines are mesons of type $s$ and $t$.
Each mass flow is handled separately, and its contribution to the $K^0$
form factor can be expressed in terms of the basic 1- and 2-loop functions
$A$, $B$, $S_{\alpha,\beta}$, $T_{\alpha_1,\alpha_2,\beta}$,
where for the latter at most the tensor indices
\begin{quote}
  $S_{2,0}$, $S_{1,1}$, $S_{1,0}$, $S_{0,0}$, $S_{0,1}$, $S_{0,2}$,
  $T_{0,0,0}$, $T_{0,0,1}$, $T_{0,0,2}$, $T_{0,0,3}$, $T_{1,0,0}$,
  $T_{1,0,1}$, $T_{1,0,2}$, $T_{1,1,0}$, $T_{1,1,1}$
\end{quote}
are needed. In the following table we give the detailed result for
each mass flow. We omit the arguments of the 2-loop functions
$S_{\alpha,\beta}$ and $T_{\alpha_1,\alpha_2,\beta}$:
for mass flow $\Diag{5b}{rst}{K^0}{}$ they are understood to be
$S_{\alpha,\beta}(m_K^2; m_r^2, m_s^2, m_t^2,)$
and $T_{\alpha_1,\alpha_2,\beta}(q^2; m_K^2,m_K^2; m_r^2,m_r^2,m_s^2,m_t^2)$.
\bea
  %~~~~~~~~~~~~~
  %  (5b), kpp
  %~~~~~~~~~~~~~
  \Diag{5b}{\Ka\pi\pi}{K^0}{} &=&
  \frac{1}{8\vierpi{4} F^4 m_\Ka^2 (4 m_\Ka^2 - q^2)} \Big\{
  \\&&\hspace{1em}
  - \; \TFull{1}{1}{1}{K}{K}{K}{K}{\pi}{\pi} \cdot 32 \: m_\Ka^2
  +    \TFull{1}{1}{0}{K}{K}{K}{K}{\pi}{\pi} \cdot 16 \: m_\Ka^2 (4 m_\Ka^2 - q^2)
  +    \TFull{1}{0}{2}{K}{K}{K}{K}{\pi}{\pi} \cdot 16 \: m_\Ka^2
  \nonumber\\&&\hspace{1em}
  + \; \TFull{1}{0}{1}{K}{K}{K}{K}{\pi}{\pi} \cdot 8 \: m_\Ka^2 ( - 4 m_\Ka^2 + q^2)
  -    \TFull{0}{0}{3}{K}{K}{K}{K}{\pi}{\pi} \cdot 2 \: m_\Ka^2
  +    \TFull{0}{0}{2}{K}{K}{K}{K}{\pi}{\pi} \cdot m_\Ka^2 (4 m_\Ka^2 - q^2)
  \nonumber\\&&\hspace{1em}
  + \; \TSun{2}{0}{K}{K}{\pi}{\pi} \cdot 16 \: (2 m_\Ka^2 - q^2)
  +    \TSun{1}{1}{K}{K}{\pi}{\pi} \cdot 8 \: ( - 4 m_\Ka^2 + q^2)
  +    \TSun{1}{0}{K}{K}{\pi}{\pi} \cdot 8 \: m_\Ka^2 (4 m_\Ka^2 - q^2)
  \nonumber\\&&\hspace{1em}
  + \; \TSun{0}{2}{K}{K}{\pi}{\pi} \cdot (6 m_\Ka^2 - q^2)
  +    \TSun{0}{1}{K}{K}{\pi}{\pi} \cdot 2 \: m_\Ka^2 ( - 4 m_\Ka^2 + q^2)
  \nonumber\\&&\hspace{1em}
  + \; A(m_\pi^2)^2 \cdot (4 m_\Ka^4 - 8 m_\Ka^2 m_\pi^2 - m_\Ka^2 q^2 + 2 m_\pi^2 q^2)
  \nonumber\\&&\hspace{1em}
  + \; A(m_\pi^2) \: \BOne{2}{1}{q^2}{m_\Ka^2}{m_\Ka^2} \cdot 4 \: m_\Ka^2 (4 m_\Ka^2 - q^2)
  \Big\} \nonumber
\\
  %~~~~~~~~~~~~~
  %  (5b), kpe
  %~~~~~~~~~~~~~
  \Diag{5b}{\Ka\pi\eta}{K^0}{} &=&
  \frac{1}{24\vierpi{4} F^4 m_\Ka^2 (4 m_\Ka^2 - q^2)} \Big\{
  \\&&\hspace{1em}
  - \; \TFull{0}{0}{3}{K}{K}{K}{K}{\pi}{\eta} \cdot 18 \: m_\Ka^2
  +    \TFull{0}{0}{2}{K}{K}{K}{K}{\pi}{\eta} \cdot 3 \: m_\Ka^2 (28 m_\Ka^2 - 3 q^2)
  +    \TFull{0}{0}{1}{K}{K}{K}{K}{\pi}{\eta} \cdot 8 \: m_\Ka^4 ( - 16 m_\Ka^2 + 3 q^2)
  \nonumber\\&&\hspace{1em}
  + \; \TFull{0}{0}{0}{K}{K}{K}{K}{\pi}{\eta} \cdot 16 \: m_\Ka^6 (4 m_\Ka^2 - q^2)
  +    \TSun{0}{2}{K}{K}{\pi}{\eta} \cdot 3 \: (10 m_\Ka^2 - q^2)
  +    \TSun{0}{1}{K}{K}{\pi}{\eta} \cdot 2 \: m_\Ka^2 ( - 44 m_\Ka^2 + 5 q^2)
  \nonumber\\&&\hspace{1em}
  + \; \TSun{0}{0}{K}{K}{\pi}{\eta} \cdot 8 \: m_\Ka^4 (8 m_\Ka^2 - q^2)
  \nonumber\\&&\hspace{1em}
  + \; A(m_\pi^2) \: A(m_\eta^2) \cdot (4 m_\Ka^4 - 8 m_\Ka^2 m_\pi^2 - m_\Ka^2 q^2 + 2 m_\pi^2 q^2)
  \nonumber\\&&\hspace{1em}
  + \; A(m_\pi^2) \: \BOne{2}{1}{q^2}{m_\Ka^2}{m_\Ka^2} \cdot 10 \: m_\Ka^2 (4 m_\Ka^2 - q^2)
  \nonumber\\&&\hspace{1em}
  + \; A(m_\eta^2) \: \BOne{2}{1}{q^2}{m_\Ka^2}{m_\Ka^2} \cdot 10 \: m_\Ka^2 (4 m_\Ka^2 - q^2)
  \Big\}
  \nonumber
\\
  %~~~~~~~~~~~~~
  %  (5b), kkk
  %~~~~~~~~~~~~~
  \Diag{5b}{\Ka\Ka\Ka}{K^0}{} &=&
  \frac{1}{12\vierpi{4} F^4 m_\Ka^2 (4 m_\Ka^2 - q^2)} \Big\{
  \\&&\hspace{1em}
  - \; \TFull{1}{1}{1}{K}{K}{K}{K}{K}{K} \cdot 24 \: m_\Ka^2
  +    \TFull{1}{1}{0}{K}{K}{K}{K}{K}{K} \cdot 12 \: m_\Ka^2 (4 m_\Ka^2 - q^2)
  +    \TFull{1}{0}{2}{K}{K}{K}{K}{K}{K} \cdot 24 \: m_\Ka^2
  \nonumber\\&&\hspace{1em}
  + \; \TFull{1}{0}{1}{K}{K}{K}{K}{K}{K} \cdot 12 \: m_\Ka^2 ( - 4 m_\Ka^2 + q^2)
  -    \TFull{0}{0}{2}{K}{K}{K}{K}{K}{K} \cdot 24 \: m_\Ka^4
  +    \TFull{0}{0}{1}{K}{K}{K}{K}{K}{K} \cdot 12 \: m_\Ka^4 (6 m_\Ka^2 - q^2)
  \nonumber\\&&\hspace{1em}
  + \; \TFull{0}{0}{0}{K}{K}{K}{K}{K}{K} \cdot 12 \: m_\Ka^6 ( - 4 m_\Ka^2 + q^2)
  +    \TSun{2}{0}{K}{K}{K}{K} \cdot 12 \: (2 m_\Ka^2 - q^2)
  +    \TSun{1}{1}{K}{K}{K}{K} \cdot 10 \: ( - 4 m_\Ka^2 + q^2)
  \nonumber\\&&\hspace{1em}
  + \; \TSun{1}{0}{K}{K}{K}{K} \cdot 8 \: m_\Ka^2 (4 m_\Ka^2 - q^2)
  +    \TSun{0}{2}{K}{K}{K}{K} \cdot (4 m_\Ka^2 - q^2)
  +    \TSun{0}{1}{K}{K}{K}{K} \cdot 24 \: m_\Ka^4
  \nonumber\\&&\hspace{1em}
  + \; \TSun{0}{0}{K}{K}{K}{K} \cdot 4 \: m_\Ka^4 ( - 10 m_\Ka^2 + q^2)
  \nonumber\\&&\hspace{1em}
  + \; A(m_\Ka^2)^2 \cdot 3 \: m_\Ka^2 ( - 4 m_\Ka^2 + q^2)
  \nonumber\\&&\hspace{1em}
  + \; A(m_\Ka^2) \: \BOne{2}{1}{q^2}{m_\Ka^2}{m_\Ka^2} \cdot 2 \: m_\Ka^2 ( - 4 m_\Ka^2 + q^2)
  \Big\}
  \nonumber
\\
  %~~~~~~~~~~~~~
  %  (5b), pkp
  %~~~~~~~~~~~~~
  \Diag{5b}{\pi \Ka\pi}{K^0}{} &=&
  \frac{1}{24\vierpi{4} F^4 m_\Ka^2 (4 m_\Ka^2 - q^2)} \Big\{
  \\&&\hspace{1em}
     \TFull{1}{1}{1}{K}{K}{\pi}{\pi}{K}{\pi} \cdot 72 \: m_\Ka^2
  +  \TFull{1}{1}{0}{K}{K}{\pi}{\pi}{K}{\pi} \cdot 36 \: m_\Ka^2 ( - 2 m_\Ka^2 - 2 m_\pi^2 + q^2)
  \nonumber\\&&\hspace{1em}
  + \; \TFull{1}{0}{2}{K}{K}{\pi}{\pi}{K}{\pi} \cdot 24 \: m_\Ka^2
  +    \TFull{1}{0}{1}{K}{K}{\pi}{\pi}{K}{\pi} \cdot 12 \: m_\Ka^2 ( - 4 m_\Ka^2 - 4 m_\pi^2 + q^2)
  \nonumber\\&&\hspace{1em}
  + \; \TFull{1}{0}{0}{K}{K}{\pi}{\pi}{K}{\pi} \cdot 12 \: m_\Ka^2 (2 m_\Ka^4 + 4 m_\Ka^2 m_\pi^2 - m_\Ka^2 q^2 + 2 m_\pi^4 - m_\pi^2 q^2)
  \nonumber\\&&\hspace{1em}
  - \; \TFull{0}{0}{3}{K}{K}{\pi}{\pi}{K}{\pi} \cdot 6 \: m_\Ka^2
  +    \TFull{0}{0}{2}{K}{K}{\pi}{\pi}{K}{\pi} \cdot 3 \: m_\Ka^2 (6 m_\Ka^2 + 6 m_\pi^2 - q^2)
  \nonumber\\&&\hspace{1em}
  + \; \TFull{0}{0}{1}{K}{K}{\pi}{\pi}{K}{\pi} \cdot 6 \: m_\Ka^2 ( - 3 m_\Ka^4 - 6 m_\Ka^2 m_\pi^2 + m_\Ka^2 q^2 - 3 m_\pi^4 + m_\pi^2 q^2)
  \nonumber\\&&\hspace{1em}
  + \; \TFull{0}{0}{0}{K}{K}{\pi}{\pi}{K}{\pi}
       \cdot 3 \: m_\Ka^2 (2 m_\Ka^2 + 2 m_\pi^2 - q^2) (m_\pi^2 + m_\Ka^2)^2
  \nonumber\\&&\hspace{1em}
  + \; \TSun{2}{0}{K}{\pi}{K}{\pi} \cdot 36 \: ( - 2 m_\Ka^2 + q^2)
  +    \TSun{1}{1}{K}{\pi}{K}{\pi} \cdot 2 \: ( - 4 m_\Ka^2 + q^2)
  \nonumber\\&&\hspace{1em}
  + \; \TSun{1}{0}{K}{\pi}{K}{\pi} \cdot 2 \: (4 m_\Ka^4 + 4 m_\Ka^2 m_\pi^2 - m_\Ka^2 q^2 - m_\pi^2 q^2)
  +    \TSun{0}{2}{K}{\pi}{K}{\pi} \cdot 2 \: (7 m_\Ka^2 - q^2)
  \nonumber\\&&\hspace{1em}
  + \; \TSun{0}{1}{K}{\pi}{K}{\pi} \cdot 4 \: ( - 7 m_\Ka^4 - 7 m_\Ka^2 m_\pi^2 + m_\Ka^2 q^2 + m_\pi^2 q^2)
  \nonumber\\&&\hspace{1em}
  + \; \TSun{0}{0}{K}{\pi}{K}{\pi} \cdot 2 \: (7 m_\Ka^2 - q^2) (m_\pi^2 + m_\Ka^2)^2
  \nonumber\\&&\hspace{1em}
  + \; A(m_\Ka^2) \: \BOne{2}{1}{q^2}{m_\pi^2}{m_\pi^2} \cdot 22 \cdot m_\Ka^2 ( - 4 m_\Ka^2 + q^2)
  \nonumber\\&&\hspace{1em}
  + \; A(m_\pi^2) \: \BOne{2}{1}{q^2}{m_\pi^2}{m_\pi^2} \cdot 8 \: m_\Ka^2 (4 m_\Ka^2 - q^2)
  \Big\}
  \nonumber
\\
  %~~~~~~~~~~~~~
  %  (5b), pke
  %~~~~~~~~~~~~~
  \Diag{5b}{\pi \Ka\eta}{K^0}{} &=&
  \frac{1}{24\vierpi{4} F^4 m_\Ka^2 (4 m_\Ka^2 - q^2)} \Big\{
  \\&&\hspace{1em}
     \TFull{1}{1}{1}{K}{K}{\pi}{\pi}{K}{\eta} \cdot 72 \: m_\Ka^2
  +  \TFull{1}{1}{0}{K}{K}{\pi}{\pi}{K}{\eta} \cdot 36 \: m_\Ka^2 ( - 2 m_\Ka^2 - 2 m_\pi^2 + q^2)
  \nonumber\\&&\hspace{1em}
  - \; \TFull{1}{0}{2}{K}{K}{\pi}{\pi}{K}{\eta} \cdot 72 \: m_\Ka^2
  +    \TFull{1}{0}{1}{K}{K}{\pi}{\pi}{K}{\eta} \cdot 12 \: m_\Ka^2 (4 m_\Ka^2 + 12 m_\pi^2 - 3 q^2)
  \nonumber\\&&\hspace{1em}
  + \; \TFull{1}{0}{0}{K}{K}{\pi}{\pi}{K}{\eta} \cdot 12 \: m_\Ka^2 (2 m_\Ka^4 - 4 m_\Ka^2 m_\pi^2 - m_\Ka^2 q^2 - 6 m_\pi^4 + 3 m_\pi^2 q^2)
  \nonumber\\&&\hspace{1em}
  + \; \TFull{0}{0}{3}{K}{K}{\pi}{\pi}{K}{\eta} \cdot 18 \: m_\Ka^2
  +    \TFull{0}{0}{2}{K}{K}{\pi}{\pi}{K}{\eta} \cdot 3 \: m_\Ka^2 ( - 2 m_\Ka^2 - 18 m_\pi^2 + 3 q^2)
  \nonumber\\&&\hspace{1em}
  + \; \TFull{0}{0}{1}{K}{K}{\pi}{\pi}{K}{\eta} \cdot 2 \: m_\Ka^2 ( - 5 m_\Ka^4 + 6 m_\Ka^2 m_\pi^2 + 3 m_\Ka^2 q^2 + 27 m_\pi^4 - 9 m_\pi^2 q^2)
  \nonumber\\&&\hspace{1em}
  + \; \TFull{0}{0}{0}{K}{K}{\pi}{\pi}{K}{\eta} \cdot
       m_\Ka^2 (q^2-2 m_\Ka^2 - 2 m_\pi^2) ( - 3 m_\pi^2 + m_\Ka^2)^2
  \nonumber\\&&\hspace{1em}
  + \; \TSun{2}{0}{K}{\pi}{K}{\eta} \cdot 36 \: ( - 2 m_\Ka^2 + q^2)
  +    \TSun{1}{1}{K}{\pi}{K}{\eta} \cdot 30 \: (4 m_\Ka^2 - q^2)
  \nonumber\\&&\hspace{1em}
  + \; \TSun{1}{0}{K}{\pi}{K}{\eta} \cdot 6 \: ( - 4 m_\Ka^4 - 20 m_\Ka^2 m_\pi^2 + m_\Ka^2 q^2 + 5 m_\pi^2 q^2)
  +    \TSun{0}{2}{K}{\pi}{K}{\eta} \cdot 6 \: ( - 7 m_\Ka^2 + q^2)
  \nonumber\\&&\hspace{1em}
  + \; \TSun{0}{1}{K}{\pi}{K}{\eta} \cdot 4 \: (m_\Ka^4 + 21 m_\Ka^2 m_\pi^2 - m_\Ka^2 q^2 - 3 m_\pi^2 q^2)
  \nonumber\\&&\hspace{1em}
  + \; \TSun{0}{0}{K}{\pi}{K}{\eta} \cdot 2 \:
       (m_\Ka^2 - 3 m_\pi^2) (3 m_\Ka^4 + 7 m_\Ka^2 m_\pi^2 - m_\Ka^2 q^2 - m_\pi^2 q^2)
  \nonumber\\&&\hspace{1em}
  + \; A(m_\Ka^2) \: A(m_\eta^2) \cdot 4 \: (12 m_\Ka^4 - 8 m_\Ka^2 m_\pi^2 - 3 m_\Ka^2 q^2 + 2 m_\pi^2 q^2)
  \nonumber\\&&\hspace{1em}
  + \; A(m_\Ka^2) \: \BOne{2}{1}{q^2}{m_\pi^2}{m_\pi^2} \cdot 2 \: m_\Ka^2 (4 m_\Ka^2 - q^2)
  \nonumber\\&&\hspace{1em}
  + \; A(m_\eta^2) \: \BOne{2}{1}{q^2}{m_\pi^2}{m_\pi^2} \cdot 16 \: m_\Ka^2 ( - 4 m_\Ka^2 + q^2)
  \Big\} \:.
\eea
Diagrams (5a) and (5c) yield the same contribution because of time
reversal invariance. In these diagrams, there are three mass flows
\begin{eqnarray}
  \Diag{5a}{}{K^0}{} &=& \Diag{5a}{\Ka\pi\pi}{K^0}{}
                       + \Diag{5a}{\Ka\pi\eta}{K^0}{}
                       + \Diag{5a}{\Ka\Ka\Ka}{K^0}{} \:,
\eea
which can be expressed by the basic functions $A$ and $S_{\alpha,\beta}$:
\bea
  %~~~~~~~~~~~~~
  %  (5ac) kpp
  %~~~~~~~~~~~~~
  \Diag{5a}{\Ka\pi\pi}{K^0}{} &=&
  \frac{1}{24\vierpi{4} F^4 m_\Ka^2} \Big\{
  - \; 8\: \TSun{2}{0}{K}{K}{\pi}{\pi}
  + \TSun{1}{1}{K}{K}{\pi}{\pi}
  + 12 \: m_\Ka^2 \: \TSun{1}{0}{K}{K}{\pi}{\pi}
  +  \TSun{0}{2}{K}{K}{\pi}{\pi}
  -  6 \: m_\Ka^2 \: \TSun{0}{1}{K}{K}{\pi}{\pi}
  \\&&\qquad\qquad
  + \; 3 \: m_\Ka^2  \: A(m_\Ka^2) \: A(m_\pi^2)
  +    2 \:  (2 m_\Ka^2 - m_\pi^2) \: A(m_\pi^2)^2
  \Big\}
  \nonumber
\\
  %~~~~~~~~~~~~~~
  %  (5ac), kpe
  %~~~~~~~~~~~~~~
  \Diag{5a}{\Ka\pi\eta}{K^0}{} &=&
  \frac{1}{48\vierpi{4} F^4 m_\Ka^2} \Big\{
  6 \: \TSun{1}{1}{K}{K}{\pi}{\eta}
  - 8 \: m_\Ka^2 \: \TSun{1}{0}{K}{K}{\pi}{\eta}
  + 3 \: \TSun{0}{2}{K}{K}{\pi}{\eta}
  + 2 \: m_\Ka^2  \:\TSun{0}{1}{K}{K}{\pi}{\eta}
  - 8 \: m_\Ka^4 \: \TSun{0}{0}{K}{K}{\pi}{\eta}
  \\&&\qquad\qquad
  - \; 6 \: m_\Ka^2 \: A(m_\Ka^2) \: A(m_\pi^2)
  -    4 \: m_\Ka^2 \: A(m_\Ka^2) \: A(m_\eta^2)
  \nonumber\\&&\qquad\qquad
  - \; 2 \: (m_\Ka^2 + m_\pi^2) \: A(m_\pi^2) \: A(m_\eta^2)
  \Big\}
  \nonumber
\\
  %~~~~~~~~~~~~~~
  %  (5ac), kkk
  %~~~~~~~~~~~~~~
  \Diag{5a}{\Ka\Ka\Ka}{K^0}{} &=&
  \frac{1}{24\vierpi{4} F^4 m_\Ka^2} \Big\{
  - 4\: \TSun{2}{0}{K}{K}{K}{K}
  + 2 \: \TSun{1}{1}{K}{K}{K}{K}
  -  \TSun{0}{2}{K}{K}{K}{K}
  + 3 \: m_\Ka^2 \: \TSun{0}{1}{K}{K}{K}{K}
  + m_\Ka^2 \: A(m_\Ka^2)^2
  \Big\} \:.
\eea

Except for special kinematic situations the genuine 2-loop integrals
$S_{\alpha,\beta}$ and $T_{\alpha_1,\alpha_2,\beta}$
cannot be calculated analytically. In appendix \ref{CalcOfIrrInt}
we describe the method how we calculated them by splitting them
into one part which contains the divergence and can be evaluated
analytically, and a second part which is finite and can be done
numerically.

%%%%%%%%%%%%%%%%%%%%%%%%%%%%%%%%%%%%%%%%%%%%%%%%%%%%%%%%%%%%%%%%%%%%%%%%%%%
%%%%%%%%%%%%%%%%%%%%%%%%%%%%%%%%%%%%%%%%%%%%%%%%%%%%%%%%%%%%%%%%%%%%%%%%%%%

\section{Evaluation of the diagrams and checks}

In the previous sections all ${\cal O}(p^6)$ contributions to the
$K^0$ form factor have been written in terms of some basic 1- and 2-loop
functions. Before evaluating the contibutions explicitly, we must specify a
renormalization scheme. 

In our calculation we are using dimensional  regularization and the so called
\emph{GL}-scheme which is defined in the following way: Each diagram of order
${\cal O}(p^{2n})$ is multiplied with a factor $e^{(1-n)\alpha(\varepsilon)}$
where $D\f{=}4\f{-}2\varepsilon$ is the dimension of space-time and 
$\alpha(\varepsilon)$ is given by
\bea\label{GL-Definition}
  (4\pi)^\varepsilon \, \Gamma(-1+\varepsilon) &=&
  -\frac{e^{\alpha(\varepsilon)}}{\varepsilon} \:,
\eea
that is
\bea
  \alpha(\varepsilon) &=& \varepsilon \!\: (1-\gamma+\log 4\pi)
  + \varepsilon^2 \Big( \frac{\pi^2}{12} + \frac{1}{2} \Big)
  + {\cal O}(\varepsilon^3) \:. \quad
\eea
Because of $\alpha(0)\f{=}0$ the total ${\cal O}(p^6)$ result is unchanged
in $D\f{=}4$ dimensions. The reason for this modification of each diagram
is to eliminate the geometric factor  $(4\pi)^\varepsilon \,
\Gamma(-1+\varepsilon)$ appearing in the 1-loop integrals $A$ and $B$.
%hhh
This renormalization scheme is very similar to the well-known
MS-bar scheme: the only difference is that in MS-bar the left-hand
side of the defining equation (\ref{GL-Definition}) is 
$\f{-}(4\pi)^\varepsilon\, \Gamma(\varepsilon)$ so that there
a geometric factor of $(4\pi)^\varepsilon \, \Gamma(\varepsilon)$
is eliminated from the 1-loop integrals.
%end hhh

The GL-scheme extends the usual 1-loop scheme introduced by Gasser
and Leutwyler \cite{GasserLeutw1}
in a natural way which can be understood from an inspection of the
renormalization  constants $L_i$ of ${\cal L}^{(4)}$:
In $D$-dimensional space-time they have dimension $D\f{-}4$ and their dimension
is generated by the mass scale $\mu$ of dimensional regularisation:
\bea
  L_i &=& \mu^{D-4} \, L_i(\mu,D) \:.
\eea
$L_i(\mu,D)$ has the same  $\mu$-dependence as a 1-loop integral, because $L_i$
itself is  independent of $\mu$. It can be expanded in a Laurent series around
$\varepsilon\f{=}0$ in the same way as a 1-loop integral:
\bea
  L_i(\mu,D) &=& \frac{L_i^{(-1)}}{\varepsilon}
  + L_i^{(0)}(\mu) + \varepsilon \, L_i^{(1)}(\mu) 
  + {\cal O}(\varepsilon^2) \:.
\eea
In the usual 1-loop scheme one chooses
\bea
  L_i^{(-1)} &=& -\frac{\Gamma_i}{32\pi^2}
  \\
  L_i^{(0)}(\mu) &=& L_i^{\mbox{\scriptsize\em ren}}(\mu)
  -\frac{\Gamma_i}{32\pi^2}\Big[1-\gamma+\log(4\pi) \Big] \:,
\eea
where $\Gamma_i$ are numbers which can be found in \cite{GasserLeutw1}.
The second term in $L_i^{(0)}$ is constructed so that it cancels in the 
$\varepsilon^0$-coefficient after multiplication with 
$e^{-\alpha(\varepsilon)}$:
\bea
  L_i^{GL}(\mu,D) &:=&
  e^{-\alpha(\varepsilon)} \!\: L_i(\mu,D)
  \;\:\:=\;\:\: 
  \frac{L_i^{(-1)}}{\varepsilon} 
  + L_i^{(0),GL}(\mu)
  + \varepsilon \, L_i^{(1),GL}(\mu) 
  + {\cal O}(\varepsilon^2)
  \label{LaurentExpansion}
\eea
with $L_i^{(0),GL}(\mu)\f{=}L_i^{\mbox{\scriptsize\em ren}}(\mu)$.

The dimension of the ${\cal L}^{(6)}$-parameter $\beta$ appearing in
(\ref{L6LagrangianRel}) can be treated in the same way:
\bea
  \beta &=& \mu^{2D-8} \: \beta(\mu,D) \:,
\eea
where $\beta(\mu,D)$ behaves like a 2-loop integral. Its Laurent
series in the above $GL$-scheme is given by
\bea
  \beta^{GL}(\mu,D) &:=&
  e^{-2\alpha(\varepsilon)} \!\: \beta(\mu,D)
  \;\:\:=\;\:\: \frac{\beta^{(-2),GL}}{\varepsilon^2}
  + \frac{\beta^{(-1),GL}(\mu)}{\varepsilon}
  + \beta^{(0),GL}(\mu) 
  + {\cal O}(\varepsilon) \:.
\eea

Before discussing the numerical results we enumerate the checks which we
performed on our calculation:
\begin{itemize}
  \item Because of charge conservation, the on-shell current matrix element in
    (\ref{DefCurrentMatrixElem}) must not contain a term  proportional to 
    $(p\f{-}p^\prime)_\mu$. This is manifestly the case for each individual
    diagram after inserting the Feynman rules into the vertices.
  \item According to the Ward-Fradkin-Takahashi identity 
    \cite{WardFradkinTakahashi} 
    the form factor must be equal to the charge of the particle for zero
    momentum transfer. This is not the case for each individual diagram, but for
    the sum of all reducible and the sum of all irreducible diagrams separately.
  \item The most important check of our calculation follows from an analysis of
    the divergent parts of the loop diagrams: their sum must be equal to the 
    negative divergent part of the tree graph (4), 
    so that the sum of all $p^6$ diagrams is
    finite. Since diagram (4) is a tree graph, it is a polynomial in masses and
    momenta and cannot produce logarithms thereof. Thus, all logarithmic terms
    must cancel in the sum of the divergent parts of the loop diagrams. For the
    irreducible diagrams, we find in the $GL$-scheme
    \bea
      \label{redContribution}
      \Delta_{\mbox{\scriptsize\em div.\ part}}^{\mbox{\scriptsize\em irr.\ loop}}
      &=&
      \frac{q^2(m_\Ka^2-m_\pi^2)}{9 (4\pi F)^4\!\; \varepsilon^2}
      + \frac{1}{432(4\pi F)^4 \!\; \varepsilon} \Big\{
      -16  m_\Ka^2 (10 m_\Ka^2-7 m_\pi^2) \: \Lg(m_\Ka^2)
      \\&&\qquad
      -m_\pi^2 (32 m_\Ka^2-80 m_\pi^2-9 q^2) \: \Lg(m_\pi^2)
      -3 q^2 (4 m_\Ka^2-m_\pi^2) \: \Lg(m_\eta^2)
      \nn\\&&\qquad
      -b_0(q^2;m_\Ka^2,m_\Ka^2) \:
         (160 m_\Ka^4-112 m_\Ka^2 m_\pi^2-76 m_\Ka^2 q^2+28 m_\pi^2 q^2+9 q^4)
      \nn\\&&\qquad
      -b_0(q^2;m_\pi^2,m_\pi^2) \:
         (32 m_\Ka^2 m_\pi^2-8 m_\Ka^2 q^2-80 m_\pi^4+56 m_\pi^2 q^2-9 q^4)
      \nn\\&&\qquad
      +80 q^2 (m_\Ka^2-m_\pi^2)
      \Big\}\nn
    \eea
    using $\Lg(m^2) = \log( m^2/(4\pi\mu^2)) + \gamma$, 
    and for the reducible 1- and 2-loop diagrams
    \bea
      \label{irrContribution}
      \Delta_{\mbox{\scriptsize\em div.\ part}}^{\mbox{\scriptsize\em red.\ loop}}
      &=&
      -\Delta_{\mbox{\scriptsize\em div.\ part}}^{\mbox{\scriptsize\em irr.\ loop}}
      + \frac{2 q^2 }{(4\pi)^2 F^4\varepsilon} (m_\Ka^2-m_\pi^2) \Lren{3} \:.
    \eea
    In fact, the logarithmic terms vanish in the sum of all diagrams, but no
    longer separately for the group of reducible and irreducible diagrams.
\end{itemize}

The previous two formulae (\ref{redContribution}) and (\ref{irrContribution}),
together with equation (\ref{diag4b}), fix the divergent part of the
relevant ${\cal L}^{(6)}$ constant $\beta$ which occurs in the $K^0$
form factor:
\bea
  \beta_{\mbox{\scriptsize\em div}} &=&
  -\frac{m_\Ka^2-m_\pi^2}{8\pi^2\varepsilon} \: L_3^{\mbox{\scriptsize\em ren}}
  \:.
\eea
Note that there is no $1/\varepsilon^2$ contribution, and as a consequence
the given $1/\varepsilon$ result is independent of the renormalization scheme
and the renormalization mass $\mu$, because it is the leading order term in the Laurent expansion of
$\beta$ w.r.t. $\varepsilon$.

For the numerical evaluation we choose the following input data:
\bea\label{Massenwerte2}
  m_\Ka &=& 495\!\;\mbox{MeV}
  \\
  m_\pi &=& 0.28 \!\: m_\Ka \;\:\:=\;\:\: 138.6\!\;\mbox{MeV}
  \\
  F_\pi &=& 92.4\!\;\mbox{MeV} \:.
\eea
Note that we have used $m_\eta^2$ as a short-hand notation for the
Gell-Mann-Okubo term $\frac{4}{3}m_K^2\f{-}\frac{1}{3}m_\pi^2$
in all contributions listed in the previous sections.
The loop integrals are calculated at mass scale $\mu\f{=}m_\rho\f{=}
770 \!\;\mbox{MeV}$. At that scale the relevant ${\cal L}^{(4)}$  parameters 
are given by
\bea
  L_3^{\mbox{\scriptsize\em ren}}(m_\rho) &=& (\f{-}35\f{\pm}13) \cdot 10^{-4}
  \\
  L_5^{\mbox{\scriptsize\em ren}}(m_\rho) &=& (14\f{\pm}5) \cdot 10^{-4}
  \\
  L_9^{\mbox{\scriptsize\em ren}}(m_\rho) &=& (69\f{\pm}7) \cdot 10^{-4} \:.
\eea
Finally, at ${\cal O}(p^6)$ not only the $\varepsilon^0$-coefficient 
$L_i^{(0),GL}\f{=}L_i^{\mbox{\scriptsize\em ren}}$
occurs in the result, but also the coefficient $L_i^{(1),GL}$ of the subsequent
order $\varepsilon^1$ of the Laurent expansion (\ref{LaurentExpansion}). 
It can be proved that these constants $L_i^{(1),GL}$ always enter the  
result in a specific combination with some ${\cal L}^{(6)}$ parameters, 
so that they are no new degrees of freedom and can be chosen arbitrarily: 
a change in their values would only lead to a redefinition of some 
${\cal L}^{(6)}$ constants. For our calculation, we choose
\bea
  L_i^{(1),GL}(m_\rho) &=& 0 \:.
\eea

With these input data and definitions we obtain the contributions shown 
in fig.~2 for the reducible and the irreducible loop diagrams. 
The corrections are to be compared to the leading order result which is 
also shown in fig.~2.

%~~~~~~~~~~~~
%  FIGURE 2
%~~~~~~~~~~~~

\begin{figure}
  \centerline{
    \includegraphics*[scale=0.9]{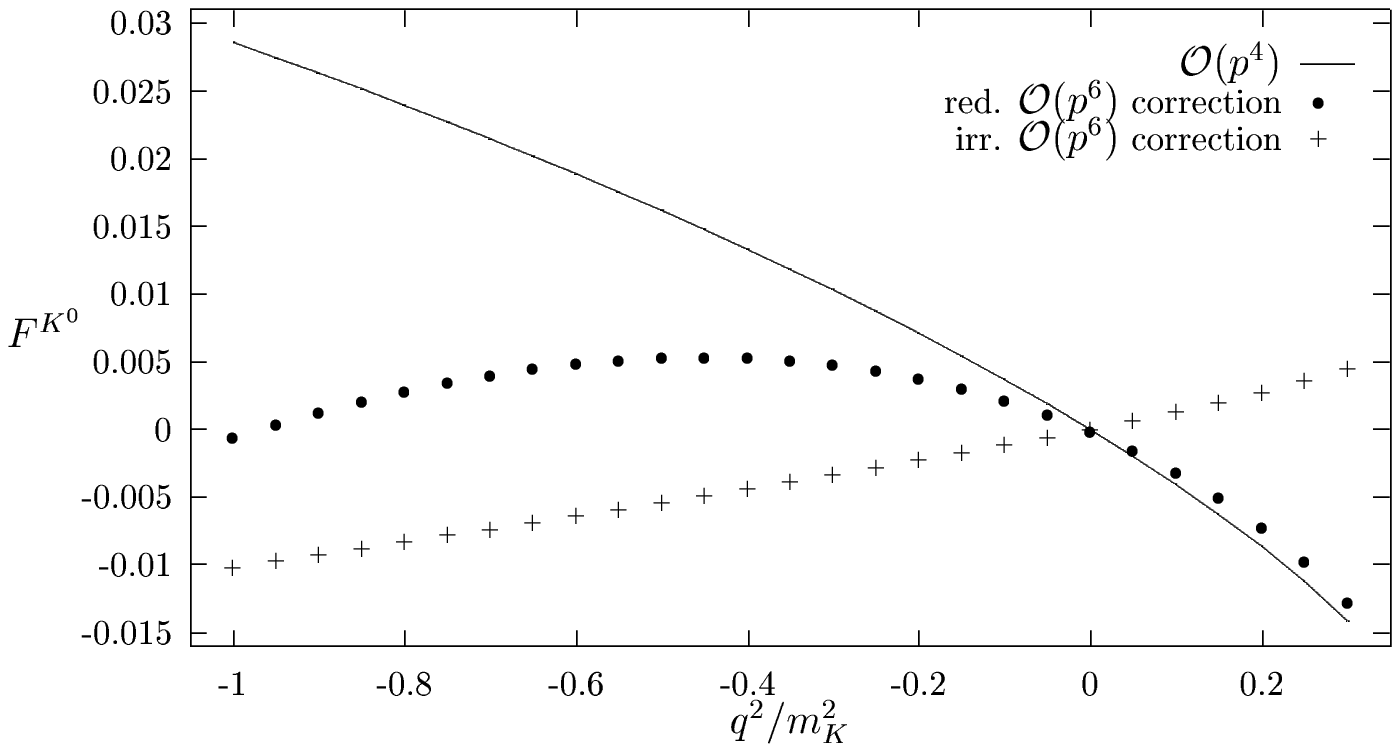}
  }
  Figure 2:
  \footnotesize
  Contributions to the $K^0$ form factor: the solid line is the leading
  order  1-loop result, the dotted curves denote the corrections at order
  $p^6$ due to the  reducible resp.\ irreducible loop diagrams in the
  $GL$-scheme (on the one hand diagrams (2a)--(3f) and the $p^6$
  contributions of the renormalized leading order diagrams (1a)--(1b), 
  on the other hand diagrams (5a)--(5c)).  The contribution of diagram (4) is
  missing: It would add a  straight line through the origin with a slope given
  by the ${\cal L}^{(6)}$-parameter $\beta$.
\end{figure}

%%%%%%%%%%%%%%%%%%%%%%%%%%%%%%%%%%%%%%%%%%%%%%%%%%%%%%%%%%%%%%%%%%%%%%%%%%%
%%%%%%%%%%%%%%%%%%%%%%%%%%%%%%%%%%%%%%%%%%%%%%%%%%%%%%%%%%%%%%%%%%%%%%%%%%%

\section{Discussion and Results}

Since at order $p^6$ there exists a direct coupling of the 
$K^{0}$ to the external photon, the slope of the form factor at zero momentum
transfer is not predicted. Diverse models make predictions for the direct
coupling $\beta $ of Eq.~(\ref{L6LagrangianRel}), but ultimately only a precise measurement of
the $K^{0}$ charge radius would fix the constant $\beta .$ On the other
hand, to extract the neutral kaon's charge radius from data a knowledge of
it's precise functional dependece on the momentum transfer $q^{2}$ is
essential. Our explicite results come here extremely handy, as only one
parameter (the charge radius) needs to be fitted to the data.

Alternatively one may consider the observable $F^{K^{0}}(q^{2})/q^{2}$ in
the comparison of theory with $\exp $eriment. Here the unknown constant 
$\beta$ only influences the vertical position of the predicted form factor
curve, but not its shape. The predictions are plotted in fig.~3. We do not
include the errors due to the ${\cal{L}}^{(4)}$ constants as their effect
tends to be proportional to $q^{2}$, and therefore cancels in the plot of
fig. 3. 

%~~~~~~~~~~~~
%  FIGURE 3
%~~~~~~~~~~~~

\begin{figure}[ht]
  \centerline{
    \includegraphics*[scale=0.9]{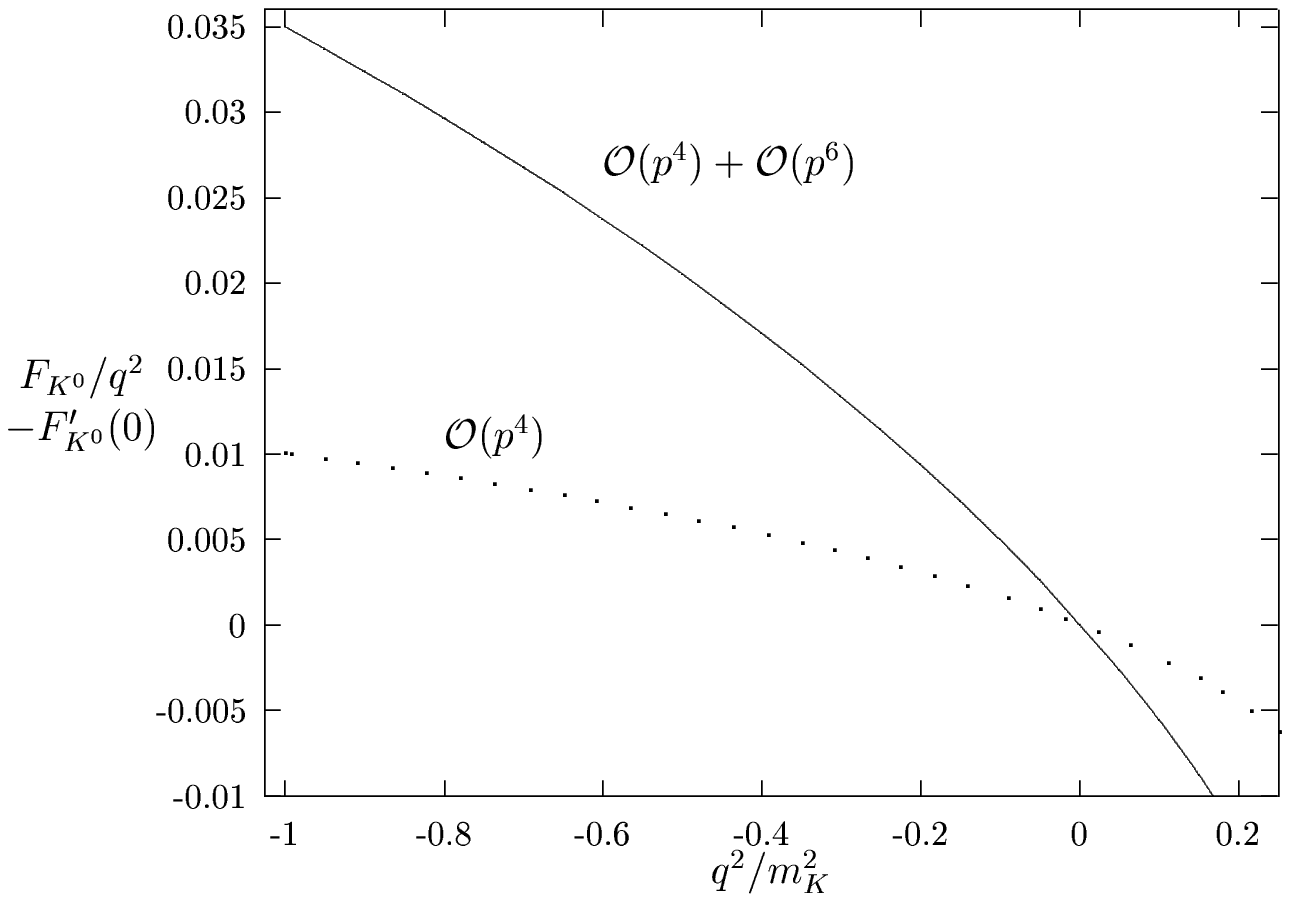}
  }
  \begin{center}
    Figure 3:
    \footnotesize
    $K^0$ form factor divided by $q^2$, in leading
    and next-to-leading order of ChPT. The curves are shifted 
    by a constant so that they go through the origin. This constant
    is not predicted by ChPT at next-to-leading order, only at
    leading order.
  \end{center}
\end{figure}

It can be observed that the ${\cal O}(p^{6})$ correction to the ${\cal O}(p^{4})$ approximation 
is unexpectedly large. This result may indicate a\  breakdown of chiral
perturbation theory or, as the ${\cal O}(p^{6})$ term represents only the first
correction to the leading ${\cal O}(p^{4})$ term, it may be due to the fact that
the ${\cal O}(p^{4})$ term is accidentally small. A detailed measurement of the
form factor of the neutal kaon may provide unique information on the
convergence of $SU(3)\times SU(3)$ chiral pertubation theory. This is
because the prediction of the shape of $F^{K^{0}}(q^{2})/q^{2}$, in contrast
to most other ${\cal O}(p^{6})$ results, does not involve any of the unknown 143
freely adjustable parameters of \cite{FearingScherer}. There exist plans for
experiments on the form factor of the neutral kaon at CEBAF and MAMY C, and
we can only hope that these experiments are realized.

%\clearpage

%%%%%%%%%%%%%%%%%%%%%%%%%%%%%%%%%%%%%%%%%%%%%%%%%%%%%%%%%%%%%%%%%%%%%%%%%%%
%%%%%%%%%%%%%%%%%%%%%%%%%%%%%%%%%%%%%%%%%%%%%%%%%%%%%%%%%%%%%%%%%%%%%%%%%%%

%%%%%%%%%%%%%%%%%%%%%
%                   %
%  A P P E N D I X  %
%                   %
%%%%%%%%%%%%%%%%%%%%%

\begin{appendix}

\section{The relevant 1-loop integrals}
\label{relevantOneLoopIntegrals}

All reducible diagrams contributing to the $K^0$~form factor
were reduced to two basic integrals in section \ref{ReducibleDiagrams}.
Here, we give the definitions and the results of these basic integrals
in dimensional regularisation.

The first basic 1-loop integral is the 1-point function
\bea
  A(m^2) &=&
  \mu^{4-D} \int \frac{d^D\!\!\:k}{i(2\pi)^D} \; \frac{1}{k^2-m^2}
  \:,
\eea
where $\mu$ is the scale of dimensional regularisation
and $D \f{=} 4 - 2\varepsilon$ is the dimension of space-time.
What is relevant in a 2-loop calculation are the coefficients of the
Laurent expansion of the integral around $\varepsilon \f{=} 0$
up to order $\varepsilon^1$. For $A$ we have
\bea
  A(m^2)
  &=&
  \frac{a_{-1}}{\varepsilon} + a_0 + \varepsilon a_1
  + {\cal O}(\varepsilon^2)
\eea
with
\bea
  a_{-1} &=& \frac{m^2}{(4\pi)^2}
  \\
  a_{0} &=& \frac{m^2}{(4\pi)^2} \Big\{ 1 - \Lg(m^2) \Big\}
  \\
  a_{1} &=& \frac{m^2}{(4\pi)^2} \Big\{ 1 + \frac{\pi^2}{12}
  - \Lg(m^2) + \frac{\Lg(m^2)^2}{2} \Big\}
\eea
where we have introduced the abbreviation
\bea
  \Lg(m^2)
  &=&
  \log\Big(\frac{m^2}{4\pi\mu^2}\Big)
  + \gamma_{\mbox{\tiny\em Euler}} \:.
\eea
The second basic 1-loop integral, the 2-point function $B(q^2,m^2)$,
is defined as the coefficient of the Lorentz-decomposition of the
tensor integral
\begin{equation}
  \mu^{4-D} \int \frac{d^D\!\!\:k}{i(2\pi)^D} \;
  \frac{k^{\mu} \, k^{\nu}}{[(k+q)^2-m^2]\:[k^2-m^2]}
\end{equation}
coming along with the metric tensor $g_{\mu\nu}$ where the other 
Lorentz-covariant is $q_\mu q_\nu$.
It can be expressed further by $A$ and the scalar 2-point function
\bea
  B_0(q^2,m^2)
  &=&
  \mu^{4-D} \int \frac{d^D\!\!\:k}{i(2\pi)^D} \;
  \frac{1}{[(k+q)^2-m^2]\:[k^2-m^2]}
\eea
in the following way: Let
\bea
  B_0(q^2,m^2) &=& \frac{b_{-1}}{\varepsilon} + b_0 + \varepsilon b_1
  + {\cal O}(\varepsilon^2)
  \\
  B(q^2,m^2) &=& \frac{(B)_{-1}}{\varepsilon} + (B)_0 + \varepsilon (B)_1
  + {\cal O}(\varepsilon^2)
\eea
be those parts of the Laurent series which are relevant in a 2-loop
calculation. $b_{-1}$, $b_0$ and $b_1$ are functions of $m^2$ and $q^2$
and are given below. 
%hhh: old text:
% The relationship between $B$ on the one side 
% and $A$ and $B_0$ on the other side is established 
% via the polynomial
% \bea
%   P(x_1,x_2,x_3) &=& \frac{1}{12} \: \Big( 
%                      x_1 + \frac{2}{3} x_2 + \frac{4}{9} x_3 
%                      \Big)\:
% \eea
% in the following way:
%end hhh: end old text
%hhh
The relationship between $B$ on the one side 
and the scalar integrals $A$ and $B_0$ on the other side follows from
a tensor decomposition:
\bea
  B(q^2,m^2)
  &=&
  \frac{1}{4(D-1)} \: \Big\{ 2 A(m^2) + (4m^2 - q^2)B_0(q^2,m^2) \Big\} \:.
\eea
The relevant Laurent coefficients of $B$ are given
in terms of those of $A$ and $B_0$ via the polynomial
\bea
  P(x_1,x_2,x_3) &=& \frac{1}{12} \: \Big( 
                     x_1 + \frac{2}{3} x_2 + \frac{4}{9} x_3 
                     \Big)\:
\eea
in the following way:
\bea\label{ReductionToScalarB0}
  (B)_{-1} &=& 2 P(a_{-1}, 0, 0) + (4m^2 - q^2) P(b_{-1}, 0, 0)
  \\
  (B)_{0} &=& 2 P(a_0, a_{-1}, 0) + (4m^2 - q^2) P(b_0, b_{-1}, 0)
  \\
  (B)_{1} &=& 2 P(a_1, a_0, a_{-1}) + (4m^2 - q^2) P(b_1, b_0, b_{-1}) \:.
\eea
The basic Laurent coefficients $b_{-1}$, $b_0$ and $b_1$ of the
scalar 2-point function $B_0$ involve logarithms and dilogarithms:
\bea
  (4\pi)^2 b_{-1} &=& 1
  \\
  (4\pi)^2 b_0 &=& 2 - \Lg(m^2) - \tau T_1
  \\
  (4\pi)^2 b_1 &=& 2 + \frac{\pi^2}{12} \, (1 - 2 \tau)
                    + \frac{1}{2} \, (\Lg(m^2) - 2)^2
                    + \; \tau \Big(
                      T_1 \Lg(m^2) - 2 T_1 - \frac{1}{2} T_1^2 + T_2 + 2 T_3
                      \Big)
\eea
where we have defined
\bea
  \tau &=& \sqrt{1 - \frac{4m^2}{q^2}}
  \\
  T_1 &=& \log \left( \frac{\tau+1}{\tau-1} \right)
  \\
  T_2 &=& \log^2 \left( \frac{\tau - 1}{2 \tau} \right)
  \\
  T_3 &=& \mbox{\rm Li}_2 \left( \frac{\tau - 1}{2 \tau} \right) \:.
\eea
An infinitesimal negative imaginary part of all masses is understood.

The Taylor expansions of the Laurent coefficients of $B$ 
w.r.t.\ $q^2$ is needed if one is interested only in the small $q^2$ 
behaviour of the form factor:
\bea
  (4\pi)^2 (B)_0 &=& - \frac{m^2}{2}  \Big\{ \Lg(m^2) - 1 \Big\}
          + \frac{q^2}{12} \: \Lg(m^2) + {\cal O}(q^4)
  \\
  (4\pi)^2 (B)_1 &=& \frac{m^2}{4} \Big\{ \big(\Lg(m^2) - 1 \big)^2 + 1 + \frac{\pi^2}{6}
          \Big\}
          - \frac{q^2}{24} \Big\{ \Lg(m^2)^2 + \frac{\pi^2}{6} \Big\}
          + {\cal O}(q^4) \:.
\eea
With the help of the above formulae the basic 1-loop functions $A$ and
$B$ and therefore all reducible loop contributions to the $K^0$ form factor
are reduced to logarithms and dilogarithms.

%%%%%%%%%%%%%%%%%%%%%%%%%%%%%%%%%%%%%%%%%%%%%%%%%%%%%%%%%%%%%%%%%%%%%%%%%%%
%%%%%%%%%%%%%%%%%%%%%%%%%%%%%%%%%%%%%%%%%%%%%%%%%%%%%%%%%%%%%%%%%%%%%%%%%%%

\section{The irreducible 2-loop integrals}
\label{CalcOfIrrInt}

In section \ref{IrreducibleDiagrams} the irreducible 2-loop diagrams 
of the $K^0$-form factor were expressed by a set of basic 2-loop
integrals of the 2-point and 3-point \emph{sunset} topologies
$S_{\alpha, \beta}$ and $T_{\alpha_1, \alpha_2, \beta}$, 
cf. (\ref{Def_von_TFull}f\/f). 
Fig.~4 shows the diagrams and the flows of momenta.

%~~~~~~~~~~~~~~~~
%  FIGURE 4
%~~~~~~~~~~~~~~~~

\begin{figure}[ht]
  \centerline{
    \includegraphics*[scale=1.0]{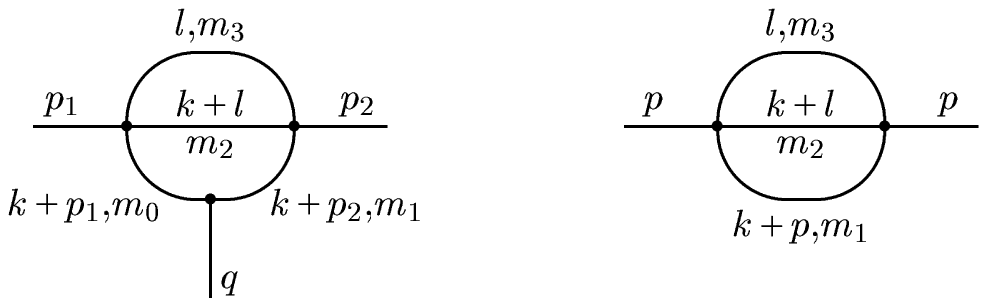}
  }
  \centerline{Figure 4}
\end{figure}

%~~~~~~~~~~~~~~~~
%  END FIGURE 4
%~~~~~~~~~~~~~~~~

In the case of three different masses ($m_\pi$, $m_\Ka$, and $m_\eta$)
it is no longer possible to reduce them to elementary analytical functions. 
Therefore, we chose a numerical approach for our calculation which 
will be outlined in the following. 
The procedure is explained in the example of the 
3-point function $T_{\alpha_1, \alpha_2, \beta}$; 
the 2-point function $S_{\alpha, \beta}$ 
is simpler and can be treated along the same lines.
Tensor integrals of the relevant topologies are decomposed into
Lorentz covariants and the algorithm is applied to the coefficient
functions.

The idea is to split the integral $T_{\alpha_1, \alpha_2, \beta}$
into two parts,
\bea
\label{decompositionOfT}
T_{\alpha_1, \alpha_2, \beta} &=& 
T_{\alpha_1, \alpha_2, \beta}^N + T_{\alpha_1, \alpha_2, \beta}^A \:,
\label{Aufspaltung} 
\eea
where $T_{\alpha_1, \alpha_2, \beta}^A$ contains the divergences, but is of
a simpler structure so that it can be calculated analytically, and
$T_{\alpha_1, \alpha_2, \beta}^N$ is finite in $D\f{=}4$ dimensions and can
be evaluated numerically.

Our method of achieving a suitable decomposition (\ref{decompositionOfT})
is based on the well-known BPHZ regularization procedure 
\cite{BPHZregularization}
which makes use of the following property of Feynman integrals:
each Feynman diagram without subdivergences behaves asymptotically
as a polynomial in its external momenta. Therefore, an UV-divergence 
of a subdivergence-free diagram may be extracted by subtracting 
a Taylor polynomial of sufficient degree w.r.t. the external momenta.

We apply this subtraction of Taylor polynomials in two steps:
First, we consider the $l$-subdiagram which is a 2-point function
with external momentum $k$, the loop momentum of the other loop integration.
Subtraction of a Taylor polynimial w.r.t. $k$ renders the $l$-subgraph 
finite and yields an integral 
\bea
\hat{T}_{\alpha_1, \alpha_2, \beta} &=& 
\int dk \: \frac{(k^2)^\alpha}{P_{k+p_1,m_0^2} P_{k+p_2,m_1^2}}
\Big( 1 - \mbox{\emph{Taylor}}_k \Big)
\int dl \: \frac{(lp_1)^{\alpha_1}(lp_2)^{\alpha_2}}{P_{k+l,m_2^2}P_{l,m_3^2}}
\eea
which has no subdivergences. The diagrammatic structure of 
$\hat{T}_{\alpha_1, \alpha_2, \beta}$ is shown in figure 5.

%~~~~~~~~~~~~~~~~
%  FIGURE 5
%~~~~~~~~~~~~~~~~

\begin{figure}[ht]
  \centerline{
    \includegraphics*[scale=1.0]{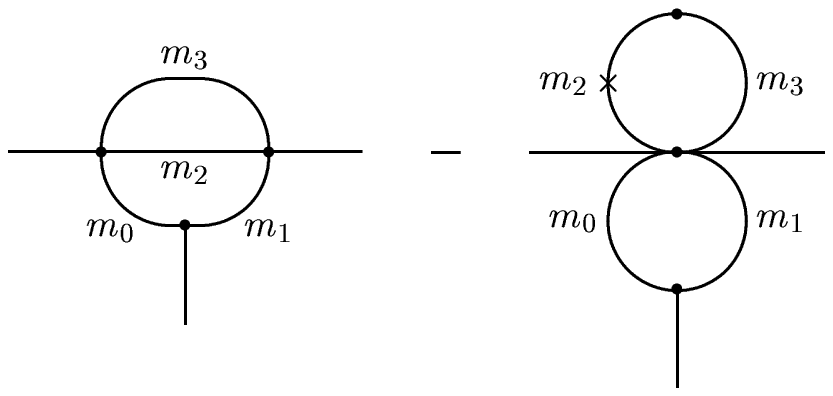}
  }
  \centerline{Figure 5 \footnotesize The propagator marked by a $\times$ sign
  is potentially taken to some integer power.}
\end{figure}

%~~~~~~~~~~~~~~~~
%  END FIGURE 5
%~~~~~~~~~~~~~~~~

In the second step we subtract a Taylor polynomial from 
$\hat{T}_{\alpha_1, \alpha_2, \beta}$ w.r.t. its external momenta
$p_1$ and $p_2$: since $\hat{T}_{\alpha_1, \alpha_2, \beta}$ has no
subdivergences left, we end up with a finite integral which is defined
to be $T_{\alpha_1, \alpha_2, \beta}^N$ in the decomposition 
(\ref{decompositionOfT}).

This decomposition has the desired properties: On the one hand, the part
$T_{\alpha_1, \alpha_2, \beta}^A$ containing the divergences has the
diagrammatic structure shown in fig.~6 and can be reduced to well-known
analytic functions. This is obvious for the first and third topology of 
fig.~6 which are products of 1-loop integrals. The second topology
in fig.~6 has the structure of a scalar 2-loop vacuum bubble
which are discussed in detail in \cite{DavTausk}.

%~~~~~~~~~~~~~~~~
%  FIGURE 6
%~~~~~~~~~~~~~~~~

\begin{figure}[ht]
  \centerline{
    \includegraphics*[scale=1.0]{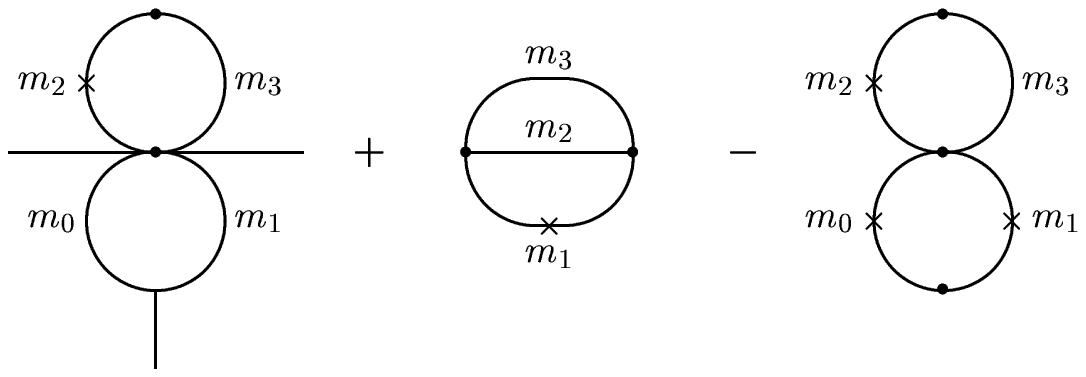}
  }
  \centerline{Figure 6 \footnotesize Propagators marked by a $\times$ sign
  are potentially taken to some integer power.}
\end{figure}

%~~~~~~~~~~~~~~~~
%  END FIGURE 6
%~~~~~~~~~~~~~~~~

On the other hand, the finite part $T_{\alpha_1, \alpha_2, \beta}^N$
is accessible to numeric evaluation, because seven of its eight integrations
can be done analytically, and the integrand of the final integration 
is a smooth function consisting of logarithms and dilogarithms.
This can be understood from the following observation: 
since the $l$-subintegral is a 2-point function with external momentum $k$,
it may be represented by a dispersion integral
\bea
  \int\limits_{(m_2+m_3)^2}^\infty d\zeta \: 
  \frac{f_{\alpha_1\alpha_2}(\zeta,m_2^2,m_3^2,p_1,p_2)}{\zeta - k^2} \:,
\eea
where the $k$ dependence occurs only in the dispersion denominator.
Interchanging the $k$ and $\zeta$ integrations we end up with an
integral
\bea
  \int\limits_{(m_2+m_3)^2}^\infty d\zeta \: 
  f_{\alpha_1\alpha_2}(\zeta,m_2^2,m_3^2,p_1,p_2)
  \int dk \: \frac{(k^2)^\beta}{P_{k+p_1,m_0^2}P_{k+p_2,m_1^2}(\zeta - k^2)} \:,
\eea
where the $k$ integration can be done analytically if one recognizes
the dispersion denominator $\zeta\f{-}k^2$ as a propagator with
mass $\zeta$ and momentum $k$. Thus, the $k$ subintegration is a 1-loop
3-point function which is given in terms of dilogarithms in 
\cite{tHooftVeltman}.

This procedure of reducing the numeric part $T_{\alpha_1, \alpha_2, \beta}^N$
to a 1-dimensional integration is illustrated diagramatically in figure 7:
writing the $l$-subgraph as a dispersion integral and subsequently
interchanging the integrations has the effect of shrinking the
$l$-subgraph to a single propagator of mass $\zeta$ which must finally
be integrated over.

%~~~~~~~~~~~~~~~~
%  FIGURE 7
%~~~~~~~~~~~~~~~~

\begin{figure}[ht]
  \centerline{
    \includegraphics*[scale=1.0]{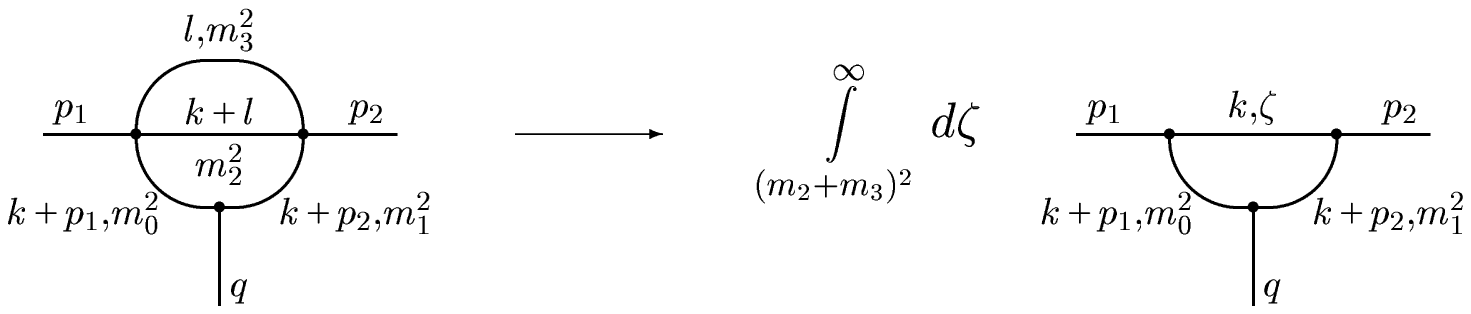}
  }
  \centerline{Figure 7} % \footnotesize 
\end{figure}

%~~~~~~~~~~~~~~~~
%  END FIGURE 7
%~~~~~~~~~~~~~~~~

The above method of calculating the irreducible 2-loop diagrams
is described in more detail in \cite{dissPeter}.
We implemented the algorithm in a set of REDUCE programs which can
be supplied on request.
 
%%%%%%%%%%%%%%%%%%%%%%%%%%%%%%%%%%%%%%%%%%%%%%%%%%%%%%%%%%%%%%%%%%%%%%%%%%%
%%%%%%%%%%%%%%%%%%%%%%%%%%%%%%%%%%%%%%%%%%%%%%%%%%%%%%%%%%%%%%%%%%%%%%%%%%%

\section{Feynman rules}
\label{FeynmanRules}

In this section we discuss the momentum space Feynman rules
which are needed for the relevant vertices occurring in the
diagrams of the $K^0$-form factor (cf. fig.~1). There are ten different
vertex types which are to be considered: vertices from
${\cal L}^{(2)}$, ${\cal L}^{(4)}$ and ${\cal L}^{(6)}$ with an even number
of meson legs and with or without an additional photon (fig.~8).

%~~~~~~~~~~~~
%  FIGURE 8
%~~~~~~~~~~~~

\begin{figure}[ht]
  \centerline{
    \includegraphics*[scale=0.9]{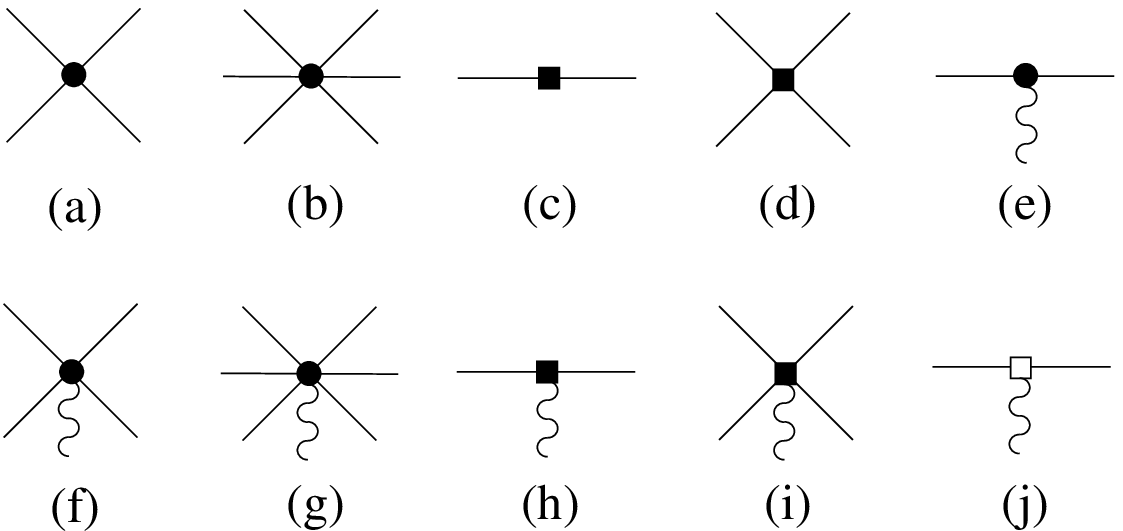}
  }
  Figure 8:
  \footnotesize
  The vertices which are needed for the $K^0$~form factor up to order $p^6$.
  A filled circle
    (\parbox{1.2ex}{ \unitlength1ex\begin{picture}(0,0)
                   \put(0.5,0){\circle*{1}}\end{picture} })
  denotes a vertex from ${\cal L}^{(2)}$, a filled square
    (\parbox{1ex}{ \unitlength1ex\begin{picture}(0,0)
                 \put(0.5,0){\makebox(0,0){\rule[0ex]{1ex}{1ex}}}
                 \end{picture} })
  a vertex from ${\cal L}^{(4)}$ and an open square
    (\parbox{1.2ex}{ \unitlength1ex\begin{picture}(0,0)
                   \put(0.5,0){\begin{picture}(0,0)\unitlength0.5ex
                     \put(-1,1){\line(1,0){2}}\put(-1,1){\line(0,-1){2}}
                     \put(1,1){\line(0,-1){2}}\put(-1,-1){\line(1,0){2}}
                   \end{picture}}\end{picture} })
  a vertex from ${\cal L}^{(6)}$.
\end{figure}

In general, the Feynman rules for a vertex
with $n$ meson legs is obtained from ${\cal L}_{\mbox{\scriptsize\em eff}}$
by first determining all monoms in the expansion of
${\cal L}_{\mbox{\scriptsize\em eff}}$ which contain exactly $n$
factors $\phi$ or derivatives thereof. Then, transformation into
momentum space is done by replacing each derivative $\partial_\mu$
by $-ip_\mu$ where $p_\mu$ is the momentum flowing \emph{into} the
corresponding leg of the vertex. Finally, the vertex must be symmetrized
in all its meson fields due to Bose symmetry.

For the vertices (a) and (b) of fig.~8 the relevant monoms of the
expansion of ${\cal L}^{(2)}$ are
\bea\label{Vierervertex}
  {\cal L}^{(2)}_{\mbox{\scriptsize\em 4-meson}} &=&
  \frac{1}{24F^2}
  \Tr\big([\phi,\partial_\mu\!\;\!\phi]\phi\:\partial^\mu\!\phi\big)
  + \frac{1}{48F^2}\:\Tr\big(\chi\phi^4\big)
  \\
  {\cal L}^{(2)}_{\mbox{\scriptsize\em 6-meson}} &=&
  \frac{1}{720F^4}\Big\{
  \Tr\big(\partial_\mu\!\;\!\phi\:\partial^\mu\!\phi\:\phi^4\big)
  -4\:\Tr\big(\partial_\mu\!\;\!\phi\:\phi\:\partial^\mu\!\phi\:\phi^3\big)
  +3\:\Tr\big(\partial_\mu\!\;\!\phi\:\phi^2\:\partial^\mu\!\phi\:\phi^2\big)
  \Big\}
  \\
  &&\qquad
  - \; \frac{1}{1440F^4}\,\Tr\big(\chi\phi^6\big)  \:. \nn
\eea
In our calculation we avoided writing down the Feynman rules for a general 
particle flow in terms of the $SU(3)$ structure constants
$f_{abc}$ and $d_{abc}$. Instead, we calculated the traces in the
Feynman rules for each concrete flow of particles separately and inserted
the Gell-Mann matrices $\lambda_a$ which occur in
\bea
  \phi &=& \sum_{a=1}^8 \phi_a \frac{\lambda_a}{2}
      \;\:\:=\;\:\:
      \left(
      \begin{array}{ccc}
      \pi^0 + \frac{1}{\sqrt{3}}\eta & \sqrt{2}\,\pi^+ & \sqrt{2} \, K^+
      \\
      \sqrt{2}\,\pi^- & -\pi^0 + \frac{1}{\sqrt{3}}\eta & \sqrt{2}\,K^0
      \\
      \sqrt{2}\, K^- & \sqrt{2} \, \bar{K}^0 & -\frac{2}{\sqrt{3}}\eta
      \end{array}\right)
\eea
directly into our REDUCE programs where also the symmetrisation is done.
In this way, the Feynman rules can be read off from the
relevant monoms of ${\cal L}_{\mbox{\scriptsize\em eff}}$. The only
step which requires some effort is the determination of these monoms.

The mass matrix can be expressed in terms of the unrenormalized meson
masses and takes the following form in the isospin limit:
\bea
  \chi &=& \mbox{\rm diag}( m_\pi^2, m_\pi^2, 2 m_K^2 - m_\pi^2 ) \:.
\eea
For vertices (c) and (d) of fig.~8 the monoms defining the Feynman rules are
\begin{eqnarray}
  %~~~~~~~~~~~~~~~~~
  %  L^4, 2-meson:
  %~~~~~~~~~~~~~~~~~
  {\cal L}^{(4)}_{\mbox{\scriptsize\em 2-meson}}
  &=&
  \frac{2 L_4}{F^2}\;\Tr\big(\partial_\mu\!\!\;\phi\:\partial^\mu\!\phi\big)\:
  \Tr\big(\chi\big)
  +
  \frac{2 L_5}{F^2}\;\Tr\big(\chi\,\partial_\mu\!\!\;\phi\:\partial^\mu\!\phi
  \big)\:
  -
  \frac{4 L_6}{F^2}\;\Tr\big(\chi\phi^2\big)\;\Tr\big(\chi\big)
  \\
  && - \;
  \frac{4 L_7}{F^2}\Big\{\Tr\big(\chi\phi\big)\Big\}^2
  -
  \frac{2 L_8}{F^2} \Big\{ \Tr\big( \chi^2\phi^2\big)
  +
  \Tr\big(\chi\phi\chi\phi\big) \Big\}
  \:,\nonumber
\\
  %~~~~~~~~~~~~~~~~~
  %  L^4, 4-meson:
  %~~~~~~~~~~~~~~~~~
  {\cal L}^{(4)}_{\mbox{\scriptsize\em 4-meson}}
  &=&
  \hphantom{+}
  \frac{ L_1}{F^4}\Big\{\Tr\big(\partial_\nu\!\!\;\phi\:\partial^\nu\!\phi\big)
  \Big\}^2
  +
  \frac{ L_2}{F^4} \;\Tr\big(\partial_\mu\!\!\;\phi\:\partial_\nu\!\;\!\phi\big)
  \;\Tr\big(\partial^\mu\!\phi\:\partial^\nu\!\phi \big)
  \\
  && + \; \frac{ L_3}{F^4} \;\Tr\big(\partial_\mu\!\!\;\phi\:\partial^\mu\!\phi
  \:\partial_\nu\!\!\;\phi\:\partial^\nu\!\phi\big)
  \nonumber
  \\
  && - \;
  \frac{ L_4}{3F^4}\Big\{
  \Tr\big([\phi,\partial_\nu\!\!\;\phi]\phi\:\partial^\nu\!\phi\big)
  \; \Tr\big(\chi\big)
  +
  3\:\Tr\big(\partial_\nu\!\!\;\phi\:\partial^\nu\!\phi\big)
  \; \Tr\big(\chi\phi^2\big)
  \Big\}
  \nonumber
  \\
  && - \;
  \frac{ L_5}{6F^4} \Big\{
  2\:\Tr\big( \chi \phi^2 \:\partial_\nu\!\!\;\phi\:\partial^\nu\!\phi \big)
  +
  3\:\Tr\big( \chi\phi \:\partial_\nu\!\!\;\phi\:\partial^\nu\!\phi \:\phi \big)
  -
  \Tr\big(\chi\phi \:\partial_\nu\!\!\;\phi \:\phi \:\partial^\nu\!\phi \big)
  \nonumber
  \\
  &&\quad\qquad
  +
  \;\Tr\big(\chi\,\partial_\nu\!\!\;\phi \:\phi^2\:\partial^\nu\!\phi \big)
  -
  \Tr\big(\chi\,\partial_\nu\!\!\;\phi \:\phi \:\partial^\nu\!\phi \:\phi \big)
  +
  2\:\Tr\big( \chi\,\partial_\nu\!\!\;\phi\:\partial^\nu\!\phi \:\phi^2 \big)
  \Big\}
  \nonumber
  \\
  && + \;
  \frac{ L_6}{3F^4} \Big\{
  \Tr\big(\chi\phi^4\big) \; \Tr\big(\chi\big)
  +3\:\big[\Tr\big( \chi\phi^2\big) \big]^2
  \Big\}
  +
  \frac{4 L_7}{3F^4}\; \Tr\big(\chi\phi^3\big) \; \Tr\big(\chi\phi\big)
 \nonumber\\
  && + \;
  \frac{ L_8}{6F^4}\Big\{
  \:\Tr\big(\chi^2\phi^4\big)
  +2\:\Tr\big(\chi\phi\chi\phi^3\big)
  +3\:\Tr\big(\chi\phi^2\chi\phi^2\big)
  +2\:\Tr\big(\chi\phi^3\chi\phi\big)
  \Big\} \:.
  \nonumber
\end{eqnarray}

The remaining vertices (e)--(j) contain an even number of mesons and
one photon line. In the following we give the corresponding Feynman
rules for an arbitrary external boson, e.g. a photon or a $W^\pm$.
The external boson enters the Lagrangian
${\cal L}_{\mbox{\scriptsize\em eff}}$ as a gauge field of the
chiral symmetry group in the covariant derivative
\begin{eqnarray}
D_\mu\!\!\;U &=& \partial_\mu\!\!\;U + i U l_\mu - i r_\mu U
\end{eqnarray}
and can be expressed by the Gell-Mann matrices $T_a = \lambda_a/2$:
\begin{equation}
  l_\mu \;=\; \sum_{a=1}^8 l_\mu^a T^a \:,
  \qquad
  r_\mu \;=\; \sum_{a=1}^8 r_\mu^a T^a \:.
\end{equation}
In case of electromagnetic interaction, the gauge boson is the photon
and is given by
\begin{equation}
  l_\mu \;=\; r_\mu
        \;=\; -eA_\mu \;\mbox{diag}(
        {\textstyle\frac{2}{3}},{\textstyle -\frac{1}{3}},{\textstyle -\frac{1}{3}}) \:.
\end{equation}
A left-handed gauge boson $l_\mu^a$ couples to the meson current
\bea
  J_{\mu,a}^L[{\cal L}_{\mbox{\scriptsize\em eff}}]
  &=&
  \frac{\delta {{\cal L}_{\mbox{\scriptsize\em eff}}}}{\delta
  l^{\mu,a}}\bigg|_{r_\mu=l_\mu=0} \:,
\eea
the analogue is true for a right handed one.
$J_{\mu,a}^{L/R}$ is therefore the relevant monomial part of
${\cal L}_{\mbox{\scriptsize\em eff}}$ which yields the Feynman rules for
the meson vertices with one external boson.
Since ${\cal L}_{\mbox{\scriptsize\em eff}}$ is symmetric under intrinsic
parity transformation, $J_{\mu,a}^L$ and $J_{\mu,a}^R$ are not independent
of each other:
\begin{equation}
  J_{\mu,a}^R(U) \;=\; J_{\mu,a}^L(U^\dagger) \quad\mbox{resp.}\quad
  J_{\mu,a}^R(\phi) \;=\; J_{\mu,a}^L(-\phi) \:.
\end{equation}
Therefore, it suffices to specify only $J_{\mu,a}^L$ for the relevant vertices.

The Feynman rule of the vertex from ${\cal L}^{(2)}$ with $n$
mesons and one external boson is given by
\bea
J^{L}_{\mu,a}[{\cal L}^{(2)}]_{\mbox{\scriptsize \em n-mesons}} &=&
- \frac{i F^2}{2} \; \frac{i^n}{n! F^n} \;
\sum_{k=0}^{n-1} {\textstyle \left( {n-1} \atop k \right)} (-1)^k
\: \Tr\big( T_a \;\! \phi^k \partial_\mu\!\!\;\phi \: \phi^{n-k-1} \big) \:.
\eea
Vertices (e)--(g) are special cases of this formula for $n=2,4,6$.

The ${\cal L}^{(4)}$-vertices with one external boson, (h) and (i),
follow from
\bea
  %~~~~~~~~~~~~~~~~~
  %  L^4, 2-meson:
  %~~~~~~~~~~~~~~~~~
  J^{L}_{\mu,a}[{\cal L}^{(4)}]_{\mbox{\scriptsize \em 2-mesons}}
  &=&
  \hphantom{+}
  \frac{2i L_4}{F^2} \; \Tr\big(T_a[\partial_\mu\!\!\;\phi,\phi]\big)
  \;\Tr\big(\chi\big)
  \\
  &&
  + \; \frac{i L_5}{F^2} \; \Tr\,T_a\big(
  \chi \;\! \partial_\mu\!\!\;\phi \, \phi
  + \partial_\mu\!\!\;\phi \:\! \chi \phi
  - \phi \chi \:\! \partial_\mu \phi
  - \phi \, \partial_\mu\!\!\;\phi \, \chi \big)
  \nonumber
  \\
  &&
  - \; \frac{i L_9}{F^2} \, \partial^\nu \;\!
  \Tr\big(T_a [\partial_\mu\!\!\;\phi,\partial_\nu\!\!\;\phi] \big)
  \nonumber
\eea
and
\bea
  %~~~~~~~~~~~~~~~~~
  %  L^4, 4-meson:
  %~~~~~~~~~~~~~~~~~
  J^{L}_{\mu,a}[{\cal L}^{(4)}]_{\mbox{\scriptsize \em 4-mesons}}
  &=&
  \hphantom{+}
  \frac{2i L_1}{F^4} \;
  \Tr\big(T_a[\partial_\mu\!\!\;\phi,\phi]\big) \;
  \Tr\big( \partial_\nu\!\!\;\phi \: \partial^\nu\!\phi \big)
  \\
  &&
  + \; \frac{2 i L_2}{F^4} \; \Tr\big(T_a [\partial^\nu\!\!\:\phi,\phi] \big) \;
  \Tr\big( \partial_\mu\!\!\;\phi \: \partial_\nu\!\!\;\phi \big)
  \nonumber
  \\
  &&
  + \; \frac{i L_3}{F^4} \; \Tr\,T_a \big(
  \{\partial_\mu\!\!\;\phi,\partial_\nu\!\!\;\phi\:\partial^\nu\!\phi\} \phi
  - \phi \{\partial_\mu\!\!\;\phi,\partial_\nu\!\!\;\phi\:\partial^\nu\!\phi\} \big)
  \nonumber
  \\
  &&
  - \; \frac{i L_4}{6 F^4} \Big\{
  6 \: \Tr\big(T_a[\partial_\mu\!\!\;\phi,\phi]\big) \; \Tr\big(\chi \phi^2\big)
  + \Tr\,T_a\big([\partial_\mu\!\!\;\phi,\phi^3]
  - 3\, \phi \:\! [\partial_\mu\!\!\;\phi,\phi]\phi \big) \; \Tr\big(\chi\big) \Big\}
  \nonumber
  \\
  &&
  - \; \frac{i L_5}{12 F^4} \Big\{ \Tr\,T_a\big(
  \chi [\partial_\mu\!\!\;\phi,\phi] \phi^2
  + 2 \, \chi \phi^2 \:\! \partial_\mu\!\!\;\phi \, \phi
  - 2 \, \phi \chi \{\partial_\mu\!\!\;\phi,\phi^2\}
  + 2 \, \partial_\mu\!\!\;\phi \, \chi \phi^3 \big)
  \nonumber\\
  && \qquad
  + \; \Tr\,T_a\big(
  4 \, \phi \chi \phi \, \partial_\mu\!\!\;\phi \, \phi
  + 3 \, [\partial_\mu\!\!\;\phi,\phi] \chi \phi^2
  + 3 \, \phi^2 \chi [\partial_\mu\!\!\;\phi,\phi]
  + 2 \, \{\partial_\mu\!\!\;\phi,\phi^2\} \chi \phi \big)
  \nonumber\\
  && \qquad
  + \; \Tr\,T_a\big(
  - 4 \, \phi \, \partial_\mu\!\!\;\phi \, \phi \chi \phi
  - 2 \, \phi^3 \chi \, \partial_\mu\!\!\;\phi
  - 2 \, \phi \, \partial_\mu\!\!\;\phi \, \phi^2 \chi
  + \phi^2 [\partial_\mu\!\!\;\phi,\phi] \chi \big)
  \Big\}
  \nonumber
  \\
  &&
  + \; \frac{i L_9}{12 F^4} \partial^\nu \Big\{ \Tr\,T_a\big(
  - 3 \, \phi [\partial_\mu\!\!\;\phi,\partial_\nu\!\!\;\phi] \phi
  + \partial_\mu\!\!\;\phi \: \phi^2 \, \partial_\nu\!\!\;\phi
  - \partial_\nu\!\!\;\phi \: \phi^2 \, \partial_\mu\!\!\;\phi \big)
  \nonumber\\
  && \qquad
  + \; \Tr\,T_a\big( 2 \, \phi^2 [\partial_\mu\!\!\;\phi,\partial_\nu\!\!\;\phi]
  + 2 \, [\partial_\mu\!\!\;\phi,\partial_\nu\!\!\;\phi] \phi^2
  - [\partial_\mu\!\!\;\phi \: \phi , \partial_\nu\!\!\;\phi \: \phi]
  - [\phi \, \partial_\mu\!\!\;\phi , \phi \, \partial_\nu\!\!\;\phi] \big)
  \Big\} \: .
  \nonumber
\eea

Finally, vertex (j) stems from ${\cal L}^{(6)}$ and takes
the following form in the notation of \cite{FearingScherer}:
\bea
  J^{L}_{\mu,a}[{\cal L}^{(6)}]_{\mbox{\scriptsize \em 2-mesons}}
  &=&
  \label{L6FeynmanRule}
  \hphantom{+}
  \frac{i B_{8}}{2 F^2} \;
  \Tr\,T_a [\chi, \{ \partial_\mu\partial_\nu\!\!\;\phi,
  \partial^\nu\!\!\:\phi \} ]
  \\
  &&
  + \; \frac{i B_{14}}{4 F^2} \; \Tr\,T_a \big(
  2 \!\: \chi \!\: \phi \, \partial_\mu\!\!\;\phi \, \chi
  - 2 \!\: \chi \, \partial_\mu\!\!\;\phi \, \phi \, \chi
  + \chi \chi \, \partial_\mu\!\!\;\phi \, \phi
  \nonumber\\
  && \hspace{6em}
  -\; \chi \chi \!\; \phi \, \partial_\mu\!\!\;\phi
  + \partial_\mu\!\!\;\phi \, \phi \!\; \chi \chi
  - \phi \, \partial_\mu\!\!\;\phi \, \chi \chi \big)
  \nonumber
  \\
  &&
  + \; \frac{i B_{15}}{2 F^2} \, \Big\{
  \Tr\big(\chi \phi \big) \;
  \Tr\,T_a [ \partial_\mu\!\!\;\phi,\chi]
  - \Tr\big(\chi \, \partial_\mu\!\!\;\phi \big) \; \Tr\,T_a [ \phi,\chi]
  \Big\}
  \nonumber
  \\
  &&
  + \; \frac{i B_{16}}{4 F^2} \; \Tr\big(\chi\big) \;
  \Tr\,T_a [ \{ \partial_\mu\!\!\;\phi,\phi\},\chi]
  \nonumber
  \\
  &&
  + \; \frac{i B_{17}}{2 F^2} \; \Tr\,T_a \big(
    \phi \, \partial_\mu\!\!\;\phi \, \chi \chi
  - \partial_\mu\!\!\;\phi \, \chi\chi \phi
  + \phi \chi\chi \, \partial_\mu\!\!\;\phi
  - \chi \chi \, \partial_\mu\!\!\;\phi \, \phi \big)
  \nonumber
  \\
  &&
  + \; \frac{i B_{18}}{2 F^2} \; \Tr\big(\chi\big) \; \Tr\,T_a \big(
    \phi \, \partial_\mu\!\!\;\phi \, \chi
  - \partial_\mu\!\!\;\phi \, \chi \,\phi
  + \phi \, \chi \, \partial_\mu\!\!\;\phi
  - \chi \, \partial_\mu\!\!\;\phi \, \phi \big)
  \nonumber
  \\
  &&
  + \; \frac{i B_{19}}{F^2} \, \Big\{
  \Tr\big(\chi \chi\big) \; \Tr\,\big( T_a [\phi,\partial_\mu\!\!\;\phi]\big)
  + 2\,\Tr(\chi[\chi,\phi]\big) \;
  \Tr\big( T_a \, \partial_\mu\!\!\;\phi\big) \Big\}
  \nonumber
  \\
  &&
  + \; \frac{i B_{20}}{F^2} \; \Tr\big(\chi \partial_\mu\!\!\;\phi \big)
  \; \Tr\,\big( T_a [\phi,\chi]\big)
  \nonumber
  \\
  &&
  + \; \frac{i B_{21}}{F^2} \; \Tr\big(\chi\big) \; \Tr\big(\chi\big)
  \; \Tr\big( T_a [\phi,\partial_\mu\!\!\;\phi] \big)
  \nonumber
  \\
  &&
  + \; \frac{i B_{22}}{F^2} \; \Tr\,T_a\big(
  \partial_\nu \partial^\nu [\partial_\mu\partial_\beta\phi,
  \partial^\beta\!\!\!\;\phi] - \partial_\mu \partial^\nu
  [\partial_\nu\partial_\beta\phi,\partial^\beta\!\!\!\;\phi] \big)
  \nonumber
  \\
  &&
  + \; \frac{i B_{23}}{F^2} \, \partial^\nu\partial^\beta \;\! \Tr\,T_a\big(
  [\partial_\mu \partial_\beta\phi,\partial_\nu\!\!\;\phi]
  -[\partial_\nu \partial_\beta\phi,\partial_\mu\!\!\;\phi] \big)
  \nonumber
  \\
  &&
  - \; \frac{i B_{24}}{F^2} \, \partial^\nu \;\! \Tr\big(
  [\partial_\mu\!\!\;\phi,\partial_\nu\!\!\;\phi] \{\chi,T_a\} \big)
  \nonumber
  \\
  &&
  - \; \frac{i B_{25}}{F^2} \, \partial^\nu \;\! \Tr\,T_a\big(
  \partial_\mu\!\!\;\phi \,\chi \;\! \partial_\nu\!\!\;\phi
  -\partial_\nu\!\!\;\phi \,\chi \;\! \partial_\mu\!\!\;\phi \big)
  \nonumber
  \\
  &&
  - \; \frac{i B_{26}}{F^2} \; \partial^\nu \;\! \Tr\big(
  T_a [\partial_\mu\!\!\;\phi,\partial_\nu\!\!\;\phi] \big) \; \Tr\big(\chi\big)
  \nonumber
  \\
  &&
  + \; \frac{i B_{27}}{2 F^2} \, \partial^\nu \;\! \Tr\,T_a\Big(
  [\partial_\mu\!\!\;\phi, \{ \partial_\nu\!\!\;\phi, \chi\}]
  - [\partial_\nu\!\!\;\phi, \{ \partial_\mu\!\!\;\phi, \chi\}] \Big)
  \:.
  \nonumber
\eea
Note that in (\ref{L6LagrangianRel}) we have given only the
part of ${\cal L}^{(6)}$ which is relevant for the $K^0$ form factor.
We convinced ourselves by inserting the Feynman rule (\ref{L6FeynmanRule})
into diagram (4) that only the terms containing $B_{24}$ and $B_{25}$
yield a nonzero contribution to the $K^0$~form factor.

\end{appendix}

%%%%%%%%%%%%%%%%%%%%%%%%%%%%%%%%%%%%%%%%%%%%%%%%%%%%%%%%%%%%%%%%%%%%%%%%%%%
%%%%%%%%%%%%%%%%%%%%%%%%%%%%%%%%%%%%%%%%%%%%%%%%%%%%%%%%%%%%%%%%%%%%%%%%%%%

\section*{Acknowledgement}
P.P.\ was supported by the ''Studienstiftung des deutschen Volkes''. We
would like to thank D.\ Broadhurst and B.\ Tausk for many interesting
discussions. 

%%%%%%%%%%%%%%%%%%%%%%%%%
%                       %
%  L I T E R A T U R E  %
%                       %
%%%%%%%%%%%%%%%%%%%%%%%%%

%%%%%%%%%%%%%%%%%%%%%%%%%%%%%%%%%%%%%%%%%%%%%%%%%%%%%%%%%%%%%%%%%%%%%%%%%%%
%%%%%%%%%%%%%%%%%%%%%%%%%%%%%%%%%%%%%%%%%%%%%%%%%%%%%%%%%%%%%%%%%%%%%%%%%%%

\end{document}